\documentclass[11pt,a4paper]{article}
\pdfoutput=1                  % for JHEP submission
\usepackage{jheppub}
\usepackage{feynmp}
\usepackage{caption}          % for subfigures
\usepackage{subcaption}       % for subfigures
\def\be{\begin{equation}}
\def\ee{\end{equation}}
\def\bea{\begin{eqnarray}}
\def\eea{\end{eqnarray}}
\def\beal{\begin{equation}\begin{aligned}}
\def\eeal{\end{aligned}\end{equation}}
\def\nn{\nonumber}

\def\braket#1{\langle #1 \rangle}

\def\Res_#1{\operatorname*{Res}_{#1}}
\def\Tr{\operatorname*{Tr}}

\def\d{\mathrm{d}}

\def\cA{\mathcal{A}}

\def\tf{\tilde{f}}
\def\shuffle{\sqcup\mathchoice{\mkern-7mu}{\mkern-7mu}{\mkern-3.2mu}{\mkern-3.8mu}\sqcup}

\def\ie{i.~e. }
\def\eg{e.~g. }
\def\etc{etc}

\def\eqn#1{Eq.~\eqref{#1}}
\def\eqns#1#2{Eqs.~\eqref{#1} and~\eqref{#2}}

\def\fig#1{Fig.~{\ref{#1}}}
\def\figs#1#2{Figs.~{\ref{#1}} and~{\ref{#2}}}

\def\Figs#1#2{Figs.~{\ref{#1}} and~{\ref{#2}}}

\def\sec#1{Section~{\ref{#1}}}

\def\app#1{Appendix~{\ref{#1}}}
\def\rcite#1{Ref.~\cite{#1}}
\def\rcites#1{Refs.~\cite{#1}}

\def\usegraph#1#2{\includegraphics[scale=1.0,trim=0 #1 0 0]{graphs/#2.pdf}}
\def\scalegraph#1#2#3{\includegraphics[scale=#1,trim=0 #2 0 0]{graphs/#3.pdf}}

\newcommand{\eqnDiag}[1]{ \vcenter{\hbox{ #1}} }
\newcommand{\eqnDelta}[1]{\vcenter{\hbox{\!\;\includegraphics[scale=1.0]{graphs/#1.pdf}\!\;}}}

\title{Full Colour for Loop Amplitudes in Yang-Mills Theory}

\author[a]{Alexander Ochirov}
\author[b]{and Ben Page}

\affiliation[a]{Higgs Centre for Theoretical Physics, School of Physics and Astronomy,\\The University of Edinburgh, Edinburgh EH9 3JZ, Scotland, UK}
\affiliation[b]{Albert-Ludwigs-Universit\"at Freiburg, Physikalisches Institut, D–79104 Freiburg, Germany}

\emailAdd{alexander.ochirov@ed.ac.uk}
\emailAdd{ben.page@physik.uni-freiburg.de}

\abstract{We present a general method to account for full colour dependence Yang-Mills amplitudes at loop level. The method fits most naturally into the framework of multi-loop integrand reduction and in a nutshell amounts to consistently retaining the colour structures of the unitarity cuts from which the integrand is gradually constructed. This technique has already been used in the recent calculation of the two-loop five-gluon amplitude in pure Yang-Mills theory with all positive helicities,
arXiv:1507.08797.
In this note, we give a careful exposition of the method and discuss its connection to loop-level Kleiss-Kuijf relations. We also explore its implications for cancellation of nontrivial symmetry factors at two loops. As an example of its generality, we show how it applies to the three-loop case in supersymmetric Yang-Mills case.}

\preprint{FR-PHENO-2016-015 \\
\phantom{~} \hfill Edinburgh 2016/20}

\begin{document}
\maketitle

%%%%%%%%%%%%%%%%%%%%%%%%%%%%%%%%%%%%%%%%%%%%%%%%%%%%
\section{Introduction}
%%%%%%%%%%%%%%%%%%%%%%%%%%%%%%%%%%%%%%%%%%%%%%%%%%%%

With the increasing precision of measurements at the Large Hadron Collider,
the accuracy of next-to-leading order (NLO) theoretical predictions
calls for improvement.
Making next-to-next-to-leading order (NNLO) computations
as standard as the NLO ones
constitutes a formidable challenge for theorists and is far from being achieved.
Indeed, all the impressive achievements beyond NLO~\cite{%NOTE: to update
Anastasiou:2000kg,Anastasiou:2000ue,Anastasiou:2001sv,Glover:2001af,%
Garland:2001tf,Garland:2002ak,Catani:2011qz,Gehrmann:2011aa,%
Czakon:2013goa,Grazzini:2013bna,Cascioli:2014yka,Gehrmann:2014fva,Chen:2014gva,%
Caola:2014iua,Czakon:2014xsa,%
Gehrmann:2015ora,Caola:2015ila,vonManteuffel:2015msa,Grazzini:2015nwa,%
Boughezal:2015dva,Boughezal:2015dra,Boughezal:2015aha,Ridder:2015dxa,%
Anastasiou:2015vya}
are so far limited to processes with four or fewer coloured external states.
Higher-multiplicity processes at NNLO
seem to demand a similar leap in computational techniques,
as already once happened at NLO, where the methods
based on (generalised) unitarity cuts~\cite{Bern:1994zx,Bern:1994cg,%
Britto:2004nc,Forde:2007mi,Britto:2005ha,Anastasiou:2006jv,Giele:2008ve}
eventually lead to the feasibility of one-loop calculations
for up to seven external partons~\cite{Bern:2013gka,Badger:2013yda}.

Extending the NLO unitarity methods to higher loop orders is a challenge. Work in the direction of addressing this problem was first made in the context of maximal unitarity~\cite{Kosower:2011ty,%
CaronHuot:2012ab,Johansson:2012zv,Johansson:2013sda,Sogaard:2013yga,%
Sogaard:2013fpa,Sogaard:2014ila,Sogaard:2014oka,Sogaard:2014jla,%
Johansson:2015ava},
which seeks to directly reconstruct the coefficients of master integrals.
An alternative approach~\cite{Mastrolia:2011pr,Badger:2012dp,Ita:2015tya}
generalises the Ossola-Papadopoulos-Pittau method
of integrand reduction~\cite{Ossola:2006us} by making use of
computational algebraic geometry~\cite{Zhang:2012ce,%
Mastrolia:2012an,Badger:2012dv,Mastrolia:2012wf,Mastrolia:2013kca}.
The $D$-dimensional version of this method has been used
to compute the most symmetric two-loop gluonic amplitudes ---
with all helicities positive (all-plus) ---
at five~\cite{Badger:2013gxa,Badger:2015lda}
and six points~\cite{Badger:2016ozq}.
Moreover, the successful integration
of the planar five-gluon amplitude~\cite{Gehrmann:2015bfy}
has led~\cite{Dunbar:2016aux,Dunbar:2016cxp}
to a direct unitarity calculation~\cite{Dunbar:2016gjb}
of the polylogarithmic part of the $n$-point two-loop all-plus amplitude
at leading colour.

In the calculation of a fully colour-dressed amplitude the number of loop-level partial amplitudes tends to grow quite fast~\cite{Bern:1997nh,Bern:2002tk,Naculich:2011ep,Edison:2011ta}. 
That is why
the full-colour integrand of the two-loop five-point all-plus amplitude
was given by
Badger, Mogull, O'Connell and one of the current authors in \rcite{Badger:2015lda} as a whole, rather than in terms of its partial amplitudes.
In this note, we expand on the colour-decomposition method
used in the calculation of \rcite{Badger:2015lda}.
The method fits into the framework
using multi-loop integrand reduction
and, in a nutshell, amounts to a consistent way of
retaining the colour factors of the generalised unitarity cuts,
from which the irreducible numerators of the integrand are extracted.
As explained in \sec{sec:Traces},
the irreducibility properties of the numerators and the top-down approach
of gradual integrand construction guarantee that no contribution
is double counted in the process.
Once all nonzero irreducible numerators are known,
the method provides a way to assemble the full-colour amplitude.

To illustrate the basic structure of the method, we begin in
\sec{sec:Traces} by applying it in the context of a trace-based colour
decomposition that is conceptually simple.
After careful consideration, one sees that this
decomposition is redundant under Kleiss-Kuif
relations~\cite{Kleiss:1988ne}. In \sec{sec:DDMbased} we
proceed to show that this redundancy can easily be eliminated by
applying the method to the colour decomposition of Del Duca, Dixon and
Maltoni (DDM)~\cite{DelDuca:1999ha,DelDuca:1999rs}, thereby resulting in shorter expressions. Further we show that at one and
two loops in this framework one can eliminate symmetry factors in a
colour-ordered calculation. In order to aid understanding, we include
in \sec{sec:DDMbased} examples of the decomposition at one, two and
three loops.
%Within the presented colour-decomposition method,
%it is natural to use Kleiss-Kuijf relations~\cite{Kleiss:1988ne},
%satisfied by tree-level amplitudes inside the unitarity cuts,
%to reduce the number of computed irreducible numerators.
%This can be done essentially by applying the colour decomposition of Del Duca, Dixon and Maltoni (DDM)~\cite{DelDuca:1999ha,DelDuca:1999rs}
%for cuts, as discussed in detail in \sec{sec:DDMbased}.
All results of this paper trivially extend to Yang-Mills theories with
supersymmetry or in higher dimensions.

%%%%%%%%%%%%%%%%%%%%%%%%%%%%%%%%%%%%%%%%%%%%%%%%%%%%
\section{Trace-based colour decomposition}
\label{sec:Traces}
%%%%%%%%%%%%%%%%%%%%%%%%%%%%%%%%%%%%%%%%%%%%%%%%%%%%

In this section we describe the colour decomposition
of a full purely gluonic amplitude at arbitrary loop order $L$,
which follows from its construction from generalised unitarity cuts
and naturally inherits the colour structures of the cuts.

In the unitarity method,
a loop amplitude is fully constrained
by a ``spanning'' set of generalised unitarity cuts~\cite{Bern:2011qt}
that are constructed out of $D$-dimensional
tree amplitudes.
The tree-level amplitudes contain both colour and kinematic information,
and a standard way to separate these in the pure-gluon case
is to reduce all the purely adjoint colour structures
that appear in the Feynman diagrams
into combinations of fundamental traces:
\be
   {\cal A}^{(0)}_n = g^{n-2} \!\!
      \sum_{\sigma\in S_n/D_n} \!\!
      T(\sigma(1),\sigma(2),\dots,\sigma(n))
      A(\sigma(1),\sigma(2),\dots,\sigma(n)) ,
\label{AtreeTraces}
\ee
where the sum is over $(n-1)!/2$ noncyclic reflection-inequivalent permutations
applied to the standard ordering and
the associated colour factors are
\beal
   T(1,2,\dots,n) & \equiv
      \Tr\!\big(T^{a_1} T^{a_2} \!\dots T^{a_n})
    + (-1)^n \Tr\!\big(T^{a_n} \!\dots T^{a_2} T^{a_1}) .
\label{TreeTraceColourFactor}
\eeal
For instance, in the three-gluon case the only element of $S_3/D_3$
is the identity permutation,
and $T(1,2,3)$ coincides with the (imaginary) structure constant
$ \tf^{a_1 a_2 a_3} = \Tr\!\big([T^{a_1},T^{a_2}]T^{a_3}\!\big) $\footnote{We
normalise the group generators to obey
$\Tr(T^{a} T^{b})=\delta^{ab}$ and $[T^a,T^b]=\tf^{abc}T^c$.}
that corresponds to $A(1,2,3)$.

%%%%%%%%% Figure %%%%%%%%%%%%%%%
\begin{figure}[t]
\centering
\includegraphics[scale=1.0]{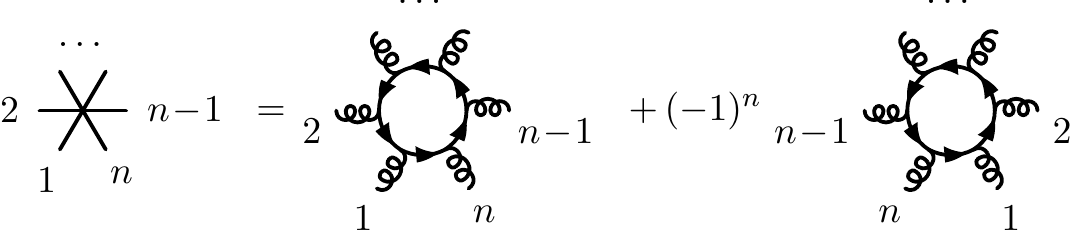}
\caption{\small Graphic version
of the definition~\eqref{TreeTraceColourFactor} for $T(1,2,\dots,n)$.
On the left-hand side it is depicted as an $n$-point vertex,
since in \sec{sec:irreducible} these combine into
trace-based colour factors for irreducible numerators.}
\label{fig:TreeTraceColourFactor}
\end{figure}
%%%%%%%%%%%%%%%%%%%%%%%%%%%%%%%%

The tree-level colour decomposition~\eqref{AtreeTraces} allows us
to write a coloured unitarity cut,
constructed out of $v$ tree amplitudes with arguments $\tau_j$,
explicitly in terms of the constituent colour-ordered amplitudes:
\be
   {\cal C}\text{ut}_{\cal I} \equiv
      \prod_{j=1}^v {\cal A}^{(0)}(\tau_j)
    = g^{n+2L-2}\!\!
      \sum_{\substack{ \sigma_1 \in S_{|\tau_1|}/D_{|\tau_1|} \\ \dots \\
                       \sigma_v \in S_{|\tau_v|}/D_{|\tau_v|}
                     }}
      \prod_{j=1}^v T(\sigma_j(\tau_j)) A(\sigma_j(\tau_j)) .
\label{CutTraces}
\ee
The labelling that we use here refers to cuts
by the sets of legs of their vertices,
which are understood in the unordered sense
for ${\cal I} = \{\tau_1,\dots,\tau_v\}$.
Now if we define
a colour-ordered cut as constructed from colour-ordered tree amplitudes,
\be
   \text{Cut}_i \equiv \prod_{j=1}^v A(\sigma_j) , \qquad \quad
   i \equiv \{\sigma_1,\dots,\sigma_v\},
\label{CutOrdered}
\ee
where its label $i$ is the ordered analogue of ${\cal I}$,
then it is natural to define the associated colour factor as
\be
   T_i \equiv \prod_{j=1}^v T(\sigma_j) , \qquad \qquad
   i = \{\sigma_1,\dots,\sigma_v\},
\label{LoopTraceColourFactor}
\ee
so that the coloured cut is simply a linear combination of 
\be
   {\cal C}\text{ut}_{\cal I}
    = g^{n+2L-2}\!\!\!
      \sum_{i=\left\{
              \substack{ \sigma_1 \in S_{|\tau_1|}/D_{|\tau_1|} \\ \dots \\
                         \sigma_v \in S_{|\tau_v|}/D_{|\tau_v|}
                       }
              \right\}}\!\!\!
      T_i\;\!\text{Cut}_i \,.
\label{CutTraces2}
\ee
In both cases, they include external particles or internal edges,
corresponding to cut propagators.
Moreover, we assume summation
over the spins and colour indices of the internal edges,\footnote{Working
at the integrand level allows us to ignore the integration
over the cut phase space in \eqn{CutTraces}.}
and this consequently applies to the cut colour factors $T_i$.

%%%%%%%%%%%%%%%%%%%%%%%%%%%%%%%%%%%%%%%%%%%%%%%%%%%%
\subsection{Cuts and irreducible numerators}
\label{sec:irreducible}
%%%%%%%%%%%%%%%%%%%%%%%%%%%%%%%%%%%%%%%%%%%%%%%%%%%%

Having defined natural colour factors for colour-ordered unitarity cuts,
we now discuss a way to translate this to the colour decomposition
of a loop amplitude.
A modern constructive way to built a multi-loop amplitude
from its unitarity cuts is the method of integrand
reduction~\cite{Ossola:2006us,Mastrolia:2011pr,Badger:2012dp,Zhang:2012ce,Mastrolia:2012an}.
It uses computational algebraic geometry~\cite{Zhang:2012ce,%
Mastrolia:2012an,Badger:2012dv,Mastrolia:2012wf,Mastrolia:2013kca}
to set up a bijection between each cut
and its corresponding irreducible numerator,
\be
   {\cal C}\text{ut}_{\cal I}
   \quad \Leftrightarrow \quad
   \tilde{\Delta}_{\cal I} ,
\label{CutNumeratorBijection}
\ee
such that the contribution of the latter to the loop integrand follows from
the condition that the loop amplitude must satisfy correct unitarity cuts:
\be
   {\cal A}^{(L)} \big|_{\cal I} = {\cal C}\text{ut}_{\cal I}
   \quad \Leftrightarrow \quad
   \int\!\frac{\d^{LD} \ell}{(2\pi)^{LD}}
         \frac{\tilde{\Delta}_{\cal I}}
              {\prod_{j\in{\cal I}} D_j} \in {\cal A}^{(L)} .
\label{CutNumeratorBijection2}
\ee

The key to this procedure is to define $\tilde{\Delta}_{\cal I}$ so that it is
equivalent to the cut on the cut kinematics, but can also be continued
off shell.  That is, unlike the cut it is obtained from,
$\tilde{\Delta}_{\cal I}$ is constructed in such a way that it can be evaluated
for arbitrary phase space points, not just those satisfying the cut
conditions. This procedure allows for a great degree of freedom and
constructing continuations which simplify final expressions
\cite{Badger:2015lda,Badger:2016ozq} or take into account IBP
relations \cite{Ita:2015tya} is an active topic of research. In full
generality, $\tilde{\Delta}_{\cal I}$ is a function which lives in the
finite-dimensional vector space of functions which, when evaluated on cut
kinematics, span the space of all numerator
functions allowed by theory power counting. Henceforth we leave the choice of basis implicit, but a
useful example to keep in mind would be independent monomials of appropriate loop variables, such as irreducible scalar products (ISPs)
of the type $(\ell_i\!\cdot\!p_j)$
that cannot be rewritten as a linear combination of the topology's propagators.

To define irreducible numerators, it is natural to start with the
maximal cuts, \ie with all the possible propagators cut.\footnote{In
$D=4-2\epsilon$ dimensions, a cut is maximal either for a trivalent
topology with $n+3(L-1)$ loop-dependent propagators or by fixing all
the $4L+L(L+1)/2$ parameters of the loop-momentum space.}  Therefore, on the maximal-cut
kinematics, the bijection~\eqref{CutNumeratorBijection} is simply
\be
   \tilde{\Delta}_{{\cal I}=\text{max.top.}}
    = {\cal C}\text{ut}_{{\cal I}=\text{max.top.}} ,
\label{TopLevel}
\ee   
from which the coordinates in the vector space are obtained.
Then $\tilde{\Delta}_{{\cal I}=\text{max.top.}}$
becomes a well-defined function in the vector
space away from the maximal-cut kinematics, despite the fact that
\eqn{TopLevel} stops being true, since the right-hand side is no longer
defined.

At this point we can see that the colour decomposition of
the maximal-topology numerator must correspond to that of the maximal cut, \ie
\be
   \tilde{\Delta}_{{\cal I}=\text{max.top.}}= g^{n+2L-2}\!\!\!
      \sum_{i=\left\{
              \substack{ \sigma_1 \in S_{|\tau_1|}/D_{|\tau_1|} \\ \dots \\
                         \sigma_v \in S_{|\tau_v|}/D_{|\tau_v|}
                       }
              \right\}}\!\!\!
      T_i\;\!\Delta_i ,
\label{TopLevelTraces}
\ee
where $\Delta_i$ are colour-ordered irreducible numerators.
The only fact that we implicitly use here is that
\emph{for different orderings $\Delta_i$
corresponding to the same $\tilde{\Delta}_{\cal I}$
we choose the same basis of numerator function space.}
In fact, we are going to adhere to this choice throughout this paper.
Moreover, the ordered topologies differing only by reflection
of one or more of its vertices are considered identical (up to a possible sign),
as implied by their reflection-inequivalent permutations sums.

Once the numerators for the maximal-level topologies are determined,
one proceeds in the top-down approach,
by gradually reducing the number of cut conditions
and thereby descending to topologies with higher-point vertices.
However, lower-level cuts are no longer polynomial functions,
since they contain poles corresponding to higher-level cuts.
In order to define the bijection~\eqref{CutNumeratorBijection}
one needs to subtract those poles,
with their residues already captured by the higher-level numerators:
\be
   \tilde{\Delta}_{\cal I}
    = {\cal C}\text{ut}_{\cal I}
    - \sum_{{\cal J}>{\cal I}}
      \frac{ \tilde{\Delta}_{\cal J} }
           { \prod_{l\in{\cal J}\setminus{\cal I}} (-i D_l) } ,
\label{LevelSubtraction}
\ee
where by ${\cal J}>{\cal I}$
we denote the topologies obtained by exposing loop-dependent propagators
concealed inside the higher-point vertices of topology ${\cal I}$.
These propagators are seen in the denominator,
with the factors of $-i$ coming from the convention
that a unitarity cut is obtained by replacing
$i/D_l\rightarrow \delta(D_l)$ in a loop amplitude.
(In practice, these can be reabsorbed into $\tilde{\Delta}_{\cal I}$.)
Thanks to this level-subtraction procedure,
the right-hand side of \eqn{LevelSubtraction},
evaluated on ${\cal C}\text{ut}_{\cal I}$-kinematics, can be used
to resolve for the basis coefficients of the function $\tilde{\Delta}_{\cal I}$.
Moreover, in this way $\tilde{\Delta}_{\cal I}/\prod_{l\in{\cal I}} D_l$
becomes the right integrand that one must add
to the previously determined part of the loop amplitude
to ensure that the latter satisfies ${\cal C}\text{ut}_{\cal I}$,
as indicated by the relationship~\eqref{CutNumeratorBijection2}.
The only subtlety here is that the cut topology ${\cal I}$
might have a nontrivial symmetry factor $S_{\cal I}$,
in general equal to the number of ways of
interchanging propagators or vertices without changing the topology.
Taking this into account, as well as that
cut diagrams have to be one-particle irreducible (1PI),
the loop amplitude can be written as a sum over such diagrams,
\be
   {\cal A}^{(L)}_n = \!\!\!
      \sum_{{\cal I}\,\in\,\text{1PI\,graphs}}
      \int\!\frac{\d^{LD} \ell}{(2\pi)^{LD}}
      \frac{\tilde{\Delta}_{\cal I}}{S_{\cal I} \prod_{l\in{\cal I}} (-i D_l)} .
\label{Acolordressed}
\ee
With this representation of the amplitude, we observe that the routing of momentum used to define the irreducible numerator is now irrelevant as the numerators live under an integral sign. We will see in \sec{sec:DDMbased} that in symmetric cases one encounters topologically equivalent terms with different loop-momentum routings. Due to the effective momentum rerouting invariance endowed by the integral sign, these terms are then seen to be equivalent, simplifying the resulting expression for the full-colour amplitude.

%%%%%%%%%%%%%%%%%%%%%%%%%%%%%%%%%%%%%%%%%%%%%%%%%%%%
\subsection{Colour ordering for irreducible numerators}
\label{sec:colourordering}
%%%%%%%%%%%%%%%%%%%%%%%%%%%%%%%%%%%%%%%%%%%%%%%%%%%%

So far we wrote the hierarchy subtraction~\eqref{LevelSubtraction}
and the resulting amplitude~\eqref{Acolordressed}
in terms of coloured objects,
but it is not hard to colour-order them.
Indeed, the hierarchy subtraction~\eqref{LevelSubtraction} is inductive
and starts with the maximal-level topologies, for which
we already have the trace-based colour decomposition~\eqref{TopLevelTraces}.
Its natural generalisation is simply
\be
   \tilde{\Delta}_{\cal I}= g^{n+2L-2}\!\!\!
      \sum_{i=\left\{
              \substack{ \sigma_1 \in S_{|\tau_1|}/D_{|\tau_1|} \\ \dots \\
                         \sigma_v \in S_{|\tau_v|}/D_{|\tau_v|}
                       }
              \right\}}\!\!\!
      T_i\;\!\Delta_i .
\label{DeltaTraces}
\ee
One can see that this is true by considering the principle of \eqn{LevelSubtraction} applied to the cut colour decomposition~\eqref{CutTraces2}. On the left-hand side of \eqref{DeltaTraces},
one subtracts all poles from ${\cal C}\text{ut}_{\cal I}$ by finding the
coloured numerators associated to more constrained cuts,
which one should here understand as factorisation limits of the original cut.
The right-hand side of \eqref{DeltaTraces}
is therefore found by considering the relevant
factorisation limits of each colour-ordered cut. In effect, \eqn{DeltaTraces} matches the colour
decomposition~\eqref{CutTraces2} of the corresponding unitarity cut
and the hierarchy subtraction~\eqref{LevelSubtraction} is then projected
onto the colour-ordered numerators:
\beal
   \Delta_i
    = \text{Cut}_i
    - \sum_{j>i}
      \frac{ \Delta_j }
           { \prod_{l \in j \setminus i} (-i D_l) } ,
\label{LevelSubtractionOrdered}
\eeal
where the ``$j\!>\!i$'' sum goes over the ordered topologies obtained
by exposing loop-dependent denominators
inside the higher-point vertices of ordered topology $i$.

Let us illustrate this projection in more detail with a short example.
Consider the following trace-basis colour decompositions
of parent and child cuts: 
\begin{align}
\label{eq:DoubleBoxColourDecomposition}
   {\cal C}\text{ut}\Big(\!\!\!\:\eqnDiag{\scalegraph{0.91}{0}{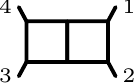}}\Big)
  & = C\Big(\!\!\!\:\eqnDiag{\scalegraph{0.91}{0}{delta331c}}\Big)
      \text{Cut}\Big(\!\!\!\:\eqnDiag{\scalegraph{0.91}{0}{delta331c}}\Big) , \\
   {\cal C}\text{ut}\Big(\!\!\!\:\eqnDiag{\scalegraph{0.91}{0}{delta231c}}\!\!\;\Big)
\label{eq:BoxTriangleColourDecomposition}
  & = C\Big(\!\!\!\:\eqnDiag{\scalegraph{0.91}{0}{delta231c}}\!\!\;\Big)
      \text{Cut}\Big(\!\!\!\:\eqnDiag{\scalegraph{0.91}{0}{delta231c}}\!\!\;\Big)
    + C\Big(\!\!\!\:\eqnDiag{\scalegraph{0.91}{0}{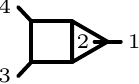}}\!\Big)
      \text{Cut}\Big(\!\!\!\:\eqnDiag{\scalegraph{0.91}{0}{delta231NPc}}\!\Big)
    \\ & \qquad \qquad \qquad \qquad \qquad \qquad~~ \nn
    + C\Big(\!\!\!\:\eqnDiag{\scalegraph{0.91}{0}{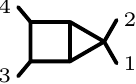}}\!\!\;\Big)
      \text{Cut}\Big(\!\!\!\:\eqnDiag{\scalegraph{0.91}{0}{delta231c2134}}\!\!\;\Big) ,
\end{align}
where the $C(\dots)$ factors are constructed
from the colour vertices~\eqref{TreeTraceColourFactor}
understood in the sense of \fig{fig:TreeTraceColourFactor}.
Notably, the two cuts are expanded onto different bases of colour structures, which obscure the irreducible numerator projection. However, taking a factorisation limit of \eqn{eq:BoxTriangleColourDecomposition} we can obtain an alternative decomposition of the parent cut that is more useful for the subtraction:
\beal
   {\cal C}\text{ut}\Big(\!\!\!\:\eqnDiag{\scalegraph{0.91}{0}{delta331c}}\Big)
  & = C\Big(\!\!\!\:\eqnDiag{\scalegraph{0.91}{0}{delta231c}}\!\!\;\Big)
      \text{Cut}\Big(\!\!\!\:\eqnDiag{\scalegraph{0.91}{0}{delta331c}}\Big)
    + C\Big(\!\!\!\:\eqnDiag{\scalegraph{0.91}{0}{delta231NPc}}\!\Big)
      \text{Cut}\Big(\!\!\!\:\eqnDiag{\scalegraph{0.91}{0}{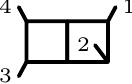}}\Big) .
%  \\
%  & = \bigg\{
%      C\Big(\!\!\!\:\eqnDiag{\scalegraph{0.91}{0}{delta231c}}\!\!\;\Big)
%    - C\Big(\!\!\!\:\eqnDiag{\scalegraph{0.91}{0}{delta231NPc}}\!\Big)
%      \bigg\}
%      \text{Cut}\Big(\!\!\!\:\eqnDiag{\scalegraph{0.91}{0}{delta331c}}\Big) .
\label{eq:AlternativeDoubleBoxColourDecomposition}
\eeal
Clearly, the parent irreducible numerator obtained from this colour expansion facilitates the colour-ordered projection~\eqref{LevelSubtractionOrdered}
applied to the child $\text{Cut}\big(\!\!\eqnDiag{\includegraphics[scale=0.67,trim=4 0 4 0,clip=true]{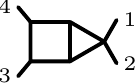}}\!\!\;\big)$.
Furthermore, here
the second ordered numerator of \eqn{eq:AlternativeDoubleBoxColourDecomposition}
is different from the one in \eqn{eq:DoubleBoxColourDecomposition}
only by a sign.
More generally, the irreducible numerators mandated by alternative parent decompositions should be obtained using KK relations~\cite{Kleiss:1988ne} from the set of irreducible numerators participating in the original parent colour decomposition, as we shall explain in more detail in \sec{sec:DDMbased}.

This example also shows that the reasoning behind the colour-ordered
projection~\eqref{LevelSubtractionOrdered} of the hierarchy
subtraction~\eqref{LevelSubtraction} is basically equivalent to that
of the tree-level colour ordering~\eqref{AtreeTraces}, which is behind
the off-shell~\cite{Berends:1987me} and
on-shell~\cite{Cachazo:2004kj,Britto:2004ap,Britto:2005fq} recursion
for tree-level objects.
Indeed, any $\Delta_j$ generated by \eqn{LevelSubtractionOrdered}
belongs to some higher-level unordered numerator $\tilde{\Delta}_{\cal J}$,
inside of which its associated colour factor is $T_j = \prod_k T(\sigma_k)$
(with repeating colour indices summed over),
just in the same way as a tree amplitude contribution
with one or more specified propagators can be related
to a product of lower-point amplitudes with similarly factorised colour factors.
Then due to the SU($N_c$)-completeness relation
\be
   T^{a}_{i \bar \jmath} T^{a}_{k \bar l}
    = \delta_{i \bar l} \delta_{k \bar \jmath}
    - \frac{1}{N_c} \delta_{i \bar \jmath} \delta_{k \bar l} ,
\label{ColourCompleteness}
\ee
products of trace-based colour factors can be mapped
to higher-point colour factors.
For instance, in the factorisation limit of an $n$-point vertex into $(k+1)$- and $(n-k+1)$-point ones, the product of two trace-based colour vertices is
\beal
   T(1,2,\dots,k,l)\,T(l,k\!+\!1,\dots,n)
    = T(1,2,\dots,k,k\!+\!1,\dots,n) & \\
    + (-1)^{k+1} T(k,\dots,2,1,k\!+\!1,\dots,n) & \\
    - \frac{1}{N_c}
      \widetilde{T}(1,2,\dots,k)
      \widetilde{T}(k\!+\!1,\dots,n) & ,
\label{TraceColourFactorProduct}
\eeal
where the $1/N_c$ term involves
$ \widetilde{T}(1,\dots,k) \equiv \Tr\!\big(T^{a_1} \!\dots T^{a_k})
                     + (-1)^{k+1} \Tr\!\big(T^{a_k} \!\dots T^{a_1}) $,
%\beal
%   \widetilde{T}(1,2,\dots,k) \equiv
%      \Tr\!\big(T^{a_1} T^{a_2} \!\dots T^{a_k})
%          + (-1)^{k+1}
%            \Tr\!\big(T^{a_k} \!\dots T^{a_2} T^{a_1}) ,
%\eeal
differing from \eqn{TreeTraceColourFactor} by the sign of the second term.
The terms with $\widetilde{T}(\dots)$ are irrelevant for the current discussion,
since they cancel altogether in the full-colour integrand due to photon-decoupling identities among $\Delta_i$,
the validity of which will be shown in \sec{sec:KK}.
The first two terms in \eqn{TraceColourFactorProduct}, however,
are colour factors of $n$-point vertices,
which thus produce poles in lower-level colour-ordered cuts
that need to be removed.

%%%%%%%%% Figure %%%%%%%%%%%%%%%
\begin{figure}[t]
\begin{subfigure}{0.37\textwidth}
\centering
\includegraphics[scale=0.82]{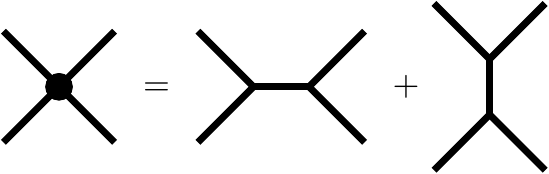}
\caption{\small\label{fig:4VertexOrdered}}
\vspace{-11pt}
\end{subfigure}
\begin{subfigure}{0.64\textwidth}
\begin{equation*}
\Rightarrow \qquad \quad
   \eqnDiag{\includegraphics[scale=1.2]{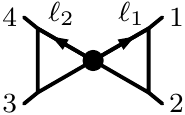}}
    = \dfrac{-i~}{(\ell_1+\ell_2)^2}
   \eqnDiag{\includegraphics[scale=1.2]{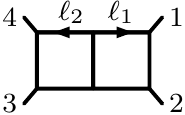}}
\end{equation*}
\vspace{-10pt}
\caption{\small\hspace{-50pt}\label{fig:Delta330Subtraction}}
\end{subfigure}
\vspace{-5pt}
\caption{\small The quartic ordered subtraction vertex contains
a maximum of two subgraphs (a).
Inside the two-loop butterfly topology (b),
only the second subgraph's propagator is loop-dependent
and contributes to the hierarchy subtraction.}
\label{fig:4VertexOrderedExample}
\end{figure}
%%%%%%%%%%%%%%%%%%%%%%%%%%%%%%%%

The resulting ordered hierarchy subtraction is by now standard in
multi-loop calculations at leading colour (see \eg \rcite{Badger:2013sta} for exposition).
Higher $n$-point vertices can in principle generate subgraphs with up to $(n-3)$
new propagators, as shown in \figs{fig:4VertexOrdered}{fig:5VertexOrdered}
(where a dotted higher-point vertex denotes the subtraction contributions
associated to that vertex).
However, since the raison d'\^etre of these subgraphs is
to remove poles from a given colour-ordered cut,
the resulting topologies that do not generate additional loop-dependent propagators do not correspond to higher-level poles.
For example, every ordered quartic vertex can give rise
to two subgraphs shown in \fig{fig:4VertexOrdered},
but in application to a two-loop numerator
$\Delta\big(\!\!\eqnDiag{\includegraphics[scale=0.67,trim=4 0 4 0,clip=true]{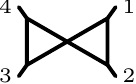}}\big)$
their denominators are $s_{12}$ and $(\ell_1+\ell_2)^2$,
so only the second subgraph actually contributes to the hierarchy subtraction,
as depicted in \fig{fig:Delta330Subtraction}.
Moreover, the number of propagators in the resulting loop graphs
must not exceed $L(L+9)/2$, as that would correspond to
an overconstrained beyond-maximal cut with non-existent pole.

%%%%%%%%% Figure %%%%%%%%%%%%%%%
\begin{figure}[t]
\centering
\includegraphics[scale=0.82]{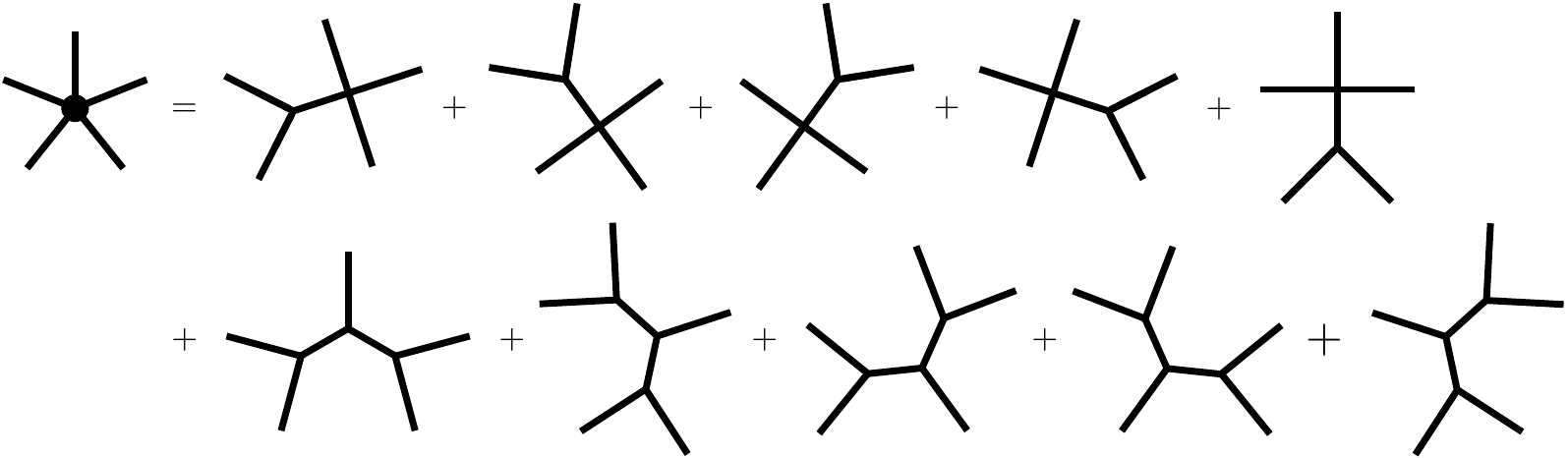}
\caption{\small The quintic ordered subtraction vertex contains a maximum
of five one-propagator and five two-propagator subgraphs.
%Only those with each propagator being loop-dependent can actually contribute to hierarchy subtraction.
These subgraphs are not Feynman diagrams of Yang-Mills theory; for instance, the six-point subtraction vertex contains even subgraphs involving five-point vertices.}
\label{fig:5VertexOrdered}
\end{figure}
%%%%%%%%%%%%%%%%%%%%%%%%%%%%%%%%

To summarise, \eqn{LevelSubtractionOrdered} defines
irreducible numerators $\Delta_i$ from colour-ordered cuts such that
they each carry a new piece of the kinematic information
invisible to the preceding more constrained cuts.
The loop amplitude~\eqref{Acolordressed} can then be consistently rewritten
as a sum over ordered one-particle irreducible graphs:\footnote{In the following
sections, the factors of $i$, inherited by \eqns{Acolordressed}{Atraces}
from the cut convention $i/D_l\rightarrow \delta(D_l)$,
will be reabsorbed into $\Delta_i$
so as to produce an overall factor of $i^{L-1}$.}
\be
   {\cal A}^{(L)}_n = g^{n+2L-2} \!\!\!
      \sum_{i\,\in\,\text{1PI\,graphs}}
      \int\!\frac{\d^{LD} \ell}{(2\pi)^{LD}}
      \frac{T_i\,\Delta_i}{S_{\cal I} \prod_{l \in i} (-i D_l)} .
\label{Atraces}
\ee
Note that since the trace-based colour factors $T_i$ are linearly dependent, this is not a decomposition into a colour basis.
Rather it is an expression given in terms of colour factors associated with
loop-singularity kinematic structures,
defined in terms of colour-ordered objects.

%%%%%%%%%%%%%%%%%%%%%%%%%%%%%%%%%%%%%%%%%%%%%%%%%%%%
\subsection{Four-point two-loop example}
\label{sec:4point2loop}
%%%%%%%%%%%%%%%%%%%%%%%%%%%%%%%%%%%%%%%%%%%%%%%%%%%%

As a simple nontrivial example of
the trace-based colour decomposition~\eqref{Atraces},
let us consider the two-loop Yang-Mills amplitude
with four plus-helicity gluons, first computed in \rcite{Bern:2000dn}.
Its helicity configuration is fully symmetric,
and its tree-level counterpart vanishes,
which is known to result in a much simpler structure
than in a generic four-point two-loop amplitude~\cite{Glover:2001af,Bern:2003ck}.
In fact its only nonzero irreducible numerators
are produced by two maximal and one next-to-maximal coloured cuts, \ie
\be
   \cA_4^{(2)}(1^+\!,2^+\!,3^+\!,4^+)
    = ig^6
      \sum_{\sigma \in S_4} \sigma \circ
      I\Bigg[ \frac{1}{4} \tilde{\Delta}\bigg(\usegraph{9}{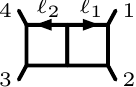}\bigg)
            + \frac{1}{4} \tilde{\Delta}\bigg(\!\usegraph{9}{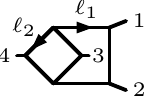}\bigg)
            + \frac{1}{8} \tilde{\Delta}\bigg(\usegraph{9}{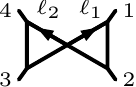}\bigg)
      \Bigg] ,
\label{AllPlus4point2loop}
\ee
where the factors of $1/4$ and $1/8$ are there
to remove overcounting of equivalent topologies in the permutation sum $S_4$.
Here and below we use a shorthand notation
for the combination of the appropriate integration measure and denominators:
\be
   I[ \tilde{\Delta}_{\cal I} ] \equiv
      \int \frac{\d^D \ell_1 \dots \d^D \ell_{L({\cal I})}}
                {(2\pi)^{DL({\cal I})}}
      \frac{\tilde{\Delta}_{\cal I}}{ \prod_{l \in {\cal I}} D_l} .
\label{IntMeasure}
\ee

The coloured irreducible numerators for the maximal topologies
are extremely simple
and depend solely on the higher-dimensional components of loop momenta,
\begin{subequations}
\begin{align}
\label{delta331}
   \tilde{\Delta}\bigg(\usegraph{9}{delta331i}\bigg)
    & = C\bigg(\usegraph{9}{delta331c}\bigg)
%        \Delta\bigg(\usegraph{9}{delta331c}\bigg)
        \frac{[12][34]}{\braket{12}\braket{34}}
        s_{12} F_1(\ell_1^{[-2\epsilon]}\!,\ell_2^{[-2\epsilon]}) , 
\\
\label{delta322}
   \tilde{\Delta}\bigg(\!\usegraph{9}{delta322i}\bigg)
    & = C\bigg(\!\usegraph{9}{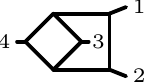}\bigg)
%        \Delta\bigg(\!\usegraph{9}{delta322c}\bigg)
        \frac{[12][34]}{\braket{12}\braket{34}}
        s_{12} F_1(\ell_1^{[-2\epsilon]}\!,\ell_2^{[-2\epsilon]}) ,
\end{align}
\label{AllPlus4point2loopDeltaMax}%
\end{subequations}
through the function
\be
   F_1(\ell_1^{[-2\epsilon]}\!,\ell_2^{[-2\epsilon]})
    = (D_s-2) \big( \mu_{11}\mu_{22}
                  + (\mu_{11}+\mu_{22})(\mu_{11}+\mu_{22}+2\mu_{12})
              \big)
    + 16 (\mu_{12}^2-\mu_{11}\mu_{22})
\label{F1}
\ee
written in terms of scalar products
$ \mu_{ij} \equiv \ell_i^{[-2\epsilon]}\!\cdot \ell_j^{[-2\epsilon]} $,
%\be
%   \mu_{ij} \equiv \ell_i^{[-2\epsilon]}\!\cdot \ell_j^{[-2\epsilon]} ,
%\label{muij}
%\ee
and the spin dimension $D_s$ that effectively interpolates between
the 't Hooft-Veltman regularisation scheme~\cite{'tHooft:1972fi},
$D_s=4-2\epsilon$,
and the four-dimensional helicity scheme~\cite{Bern:2002zk}, $D_s=4$.

The coloured numerator for the next-to-maximal topology
contains a four-point vertex with three possible colour orderings,
\beal
   \tilde{\Delta}\bigg(\usegraph{9}{delta330i}\bigg)
    = C\bigg(\usegraph{9}{delta330c}\bigg)
      \Delta\bigg(\usegraph{9}{delta330i}\bigg) &
    + C\bigg(\usegraph{9}{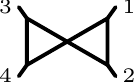}\bigg)
      \Delta\bigg(\usegraph{10}{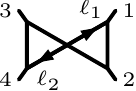}\bigg) \\ &
    + C\bigg(\usegraph{9}{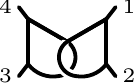}\bigg)
      \Delta\bigg(\usegraph{9}{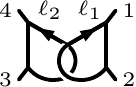}\bigg) ,
\label{delta330}
\eeal
where the colour factors $C_i$ are obtained
by plugging in the trace-based vertices~\eqref{TreeTraceColourFactor},
\begin{subequations}
\begin{align}
\label{C330a}
   C\Big(\!\!\!\:\eqnDiag{\scalegraph{0.91}{0}{delta330c}}\Big) & =
%   C\bigg(\usegraph{9}{delta330c}\bigg) & =
      T(1,e,a) T(2,b,e) T(3,f,c) T(4,d,f) T(a,b,c,d) , \\
\label{C330b}
   C\Big(\!\!\!\:\eqnDiag{\scalegraph{0.91}{0}{delta330c1243}}\Big) & =
%   C\bigg(\usegraph{9}{delta330c1243}\bigg) & =
      T(1,e,a) T(2,b,e) T(3,f,c) T(4,d,f) T(a,b,d,c) , \\
\label{C330NPa}
   C\Big(\!\!\!\:\eqnDiag{\scalegraph{0.91}{0}{delta330NPc}}\Big) & =
%   C\bigg(\usegraph{9}{delta330NPc}\bigg) & =
      T(1,e,a) T(2,b,e) T(3,f,c) T(4,d,f) T(a,c,b,d) .
\end{align}
\label{C330}%
\end{subequations}
Whilst we now have the full decomposition in the trace basis, this
expression can be further simplified. To demonstrate this, consider
the form of the ordered numerators $\Delta_i$~\cite{Bern:2000dn,Badger:2013gxa}:
\small
\begin{subequations}
\begin{align}
\label{delta330a}
   \Delta\bigg(\usegraph{9}{delta330i}\bigg)
    & = \frac{[12][34]}{\braket{12}\braket{34}}
        \bigg\{ 4(D_s\!-\!2)(\mu_{11}+\mu_{22})\mu_{12}  
%                \\ & \qquad \qquad \qquad\,
              + (D_s\!-\!2)^2 \mu_{11}\mu_{22}
                \frac{(\ell_1+\ell_2)^2+s_{12}}{s_{12}}
        \bigg\} , \\
\label{delta330b}
   \Delta\bigg(\usegraph{10}{delta330i1243}\bigg)
    & = \frac{[12][34]}{\braket{12}\braket{34}}
        \bigg\{\!-4(D_s\!-\!2)(\mu_{11}+\mu_{22})\mu_{12}
%                \\ & \qquad \qquad \qquad\,
              + (D_s\!-\!2)^2 \mu_{11}\mu_{22}
                \frac{(\ell_1-\ell_2-p_{12})^2+s_{12}}{s_{12}}
        \bigg\} , \\
\label{delta330NP}
   \Delta\bigg(\usegraph{9}{delta330NPi}\bigg)
    & =-\frac{[12][34]}{\braket{12}\braket{34}}
        (D_s\!-\!2)^2 \mu_{11}\mu_{22}
        \frac{(\ell_1+\ell_2)^2 +(\ell_1-\ell_2-p_{12})^2+2s_{12}}{s_{12}} .
\end{align}
\label{AllPlus4point2loopDeltaNMax}%
\end{subequations}
\normalsize

We can immediately observe that they obey
\be
   \Delta\bigg(\usegraph{9}{delta330i}\bigg)
 + \Delta\bigg(\usegraph{9}{delta330i1243}\bigg)
 + \Delta\bigg(\usegraph{9}{delta330NPi}\bigg) = 0 .
\label{KK330}
\ee
We dub this a numerator KK relation in that it is inherited 
from the same relation between the colour-ordered cuts.
To understand how this arises note that the three corresponding
cuts contain identical three-point amplitudes and differ only by the
orderings of the four-point amplitude in the middle, which naturally
obey the KK relation~\cite{Kleiss:1988ne}
\be
   A(\ell_1,\ell_2,\ell_3,\ell_4)
 + A(\ell_1,\ell_3,\ell_2,\ell_4)
 + A(\ell_1,\ell_2,\ell_4,\ell_3) = 0 ,
\label{KK4point}
\ee
where $\ell_3=-\ell_2-p_{12}$, $\ell_4=-\ell_1+p_{12}$.
Therefore,
\be
   \text{Cut}\Bigg(\usegraph{13}{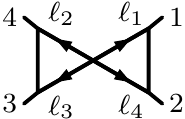}\Bigg)
 + \text{Cut}\Bigg(\usegraph{13}{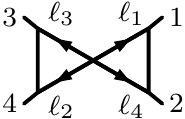}\Bigg)
 + \text{Cut}\Bigg(\usegraph{13}{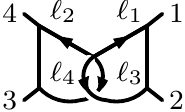}\Bigg) = 0 .
\label{KK330cuts}
\ee
Each of these cuts is given by its irreducible numerator
and the higher-level numerator from the maximal topology,
as indicated in \fig{fig:4VertexOrderedExample},
but the latter cancel in the sum:
\beal
 & \Delta\bigg(\usegraph{9}{delta330i}\bigg)
 + \frac{1}{(\ell_1+\ell_2)^2}
   \Delta\bigg(\usegraph{9}{delta331i}\bigg) \\ &
 + \Delta\bigg(\usegraph{9}{delta330i1243}\bigg)
 + \frac{1}{(\ell_1-\ell_2-p_{12})^2}
   \Delta\bigg(\usegraph{9}{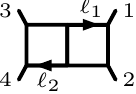}\bigg) \\ &
 + \Delta\bigg(\usegraph{9}{delta330NPi}\bigg)
 - \frac{1}{(\ell_1+\ell_2)^2}
   \Delta\bigg(\usegraph{9}{delta331i}\bigg)
 - \frac{1}{(\ell_1-\ell_2-p_{12})^2}
   \Delta\bigg(\usegraph{9}{delta331i1243}\bigg) = 0 ,
\label{KK330verbose}
\eeal
yielding exactly \eqn{KK330} on the cut.

To ensure that it holds away from the cut kinematics, we rely on
the bijection properties of the cut-numerator map~\eqref{CutNumeratorBijection}.
In other words, the loop variables chosen for
the three orderings~\eqref{delta330a}, \eqref{delta330b} and~\eqref{delta330NP}
should be consistent with a single choice of the irreducible monomials
for the butterfly topology in the unordered sense.
At first glance, it may seem not to be the case,
as on the cut kinematics among the three appearing variables 
\beal
   s_{12} , \qquad \quad
   (\ell_1+\ell_2)^2 = 2\ell_1\!\cdot\!\ell_2 , \qquad \quad
   (\ell_1-\ell_2-p_{12})^2 = -2\ell_1\!\cdot\!\ell_2-s_{12} ,
\eeal
only any two can be considered as independent monomials
(along with the factor $\mu_{11}\mu_{22}$).
However, there is only two unique combinations that appear
in the numerators~\eqn{AllPlus4point2loopDeltaNMax}:
\beal
   (\ell_1+\ell_2)^2+s_{12} = 2\ell_1\!\cdot\!\ell_2+s_{12} , \qquad \quad
   (\ell_1-\ell_2-p_{12})^2+s_{12} = -2\ell_1\!\cdot\!\ell_2 ,
\label{monomialsubset330}
\eeal
that do form an independent monomial set, which is equivalent to
$\{\mu_{11}\mu_{22},\mu_{11}\mu_{22}(\ell_1\!\cdot\!\ell_2)\}$.
The choice of the set~\eqref{monomialsubset330}
is dictated by the empirical knowledge
that the inclusion of harmless $\ell_1^2$, $\ell_2^2$, \etc.,
inside the monomials results in vanishing lower-level numerators.

In \sec{sec:KK} we will give a more general argument for the validity
of the KK relations~\cite{Kleiss:1988ne} between irreducible numerators.
Now let us proceed with taking advantage of \eqn{KK330}.
Plugging it in the coloured numerator~\eqref{delta330}, we obtain
\beal
   \tilde{\Delta}\bigg(\usegraph{9}{delta330i}\bigg)
    =~&\bigg\{
      C\Big(\!\!\!\:\eqnDiag{\scalegraph{0.91}{0}{delta330c}}\Big)
    - C\Big(\!\!\!\:\eqnDiag{\scalegraph{0.91}{0}{delta330NPc}}\Big)
      \bigg\}
      \Delta\bigg(\usegraph{9}{delta330i}\bigg) \\
    + &\bigg\{
      C\Big(\!\!\!\:\eqnDiag{\scalegraph{0.91}{0}{delta330c1243}}\Big)
    - C\Big(\!\!\!\:\eqnDiag{\scalegraph{0.91}{0}{delta330NPc}}\Big)
      \bigg\}
      \Delta\bigg(\usegraph{10}{delta330i1243}\bigg) \\
    =~&C\bigg(\usegraph{9}{delta331c}\bigg)
      \Delta\bigg(\usegraph{9}{delta330i}\bigg)
    + C\bigg(\usegraph{9}{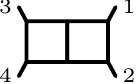}\bigg)
      \Delta\bigg(\usegraph{10}{delta330i1243}\bigg) ,
\label{delta330DDM}
\eeal
where in the last line we used that the colour traces in \eqn{C330} satisfy
\be
   T(a,b,c,d) - T(a,c,b,d) = \tf^{bce} \tf^{eda} , \qquad \quad
   T(a,b,d,c) - T(a,c,b,d) = \tf^{bde} \tf^{eca} ,
\label{ColourReductionReverse}
\ee
\ie we reversed the familiar colour reduction for the four-point cubic diagrams.
Comparing \eqns{delta330}{delta330DDM},
note that the two colour decompositions of the butterfly numerator
are obtained respectively from the two expressions for the four-point tree,
\beal
   {\cal A}^{(0)}_4 & = T(1,2,3,4) A(1,2,3,4)
                      + T(1,2,4,3) A(1,2,4,3)
                      + T(1,3,2,4) A(1,3,2,4) \\
    & = \tf^{23e} \tf^{e41} A(1,2,3,4) + \tf^{24e} \tf^{e31} A(1,2,4,3) ,
\label{AtreeDDM4point}
\eeal
the latter being the DDM decomposition~\cite{DelDuca:1999rs}.

Since we have thus eliminated
the colour-ordered butterfly numerator~\eqref{delta330NP},
and the numerators~\eqref{delta330a} and~\eqref{delta330b}
are the same up to relabelling,
we can rewrite the full amplitude~\eqref{AllPlus4point2loop}
in a compact form:
\beal \!\!\!\!\!
   \cA_4^{(2)}(1^+\!,2^+\!,3^+\!,4^+)
    = \frac{ig^6}{4}
      \sum_{\sigma \in S_4} \sigma \circ
      I\Bigg[\,
              C\bigg(\usegraph{9}{delta331c}\bigg)
              \Bigg( &
              \Delta\bigg(\usegraph{9}{delta331i}\bigg)
            + \Delta\bigg(\usegraph{9}{delta330i}\bigg)\!
              \Bigg) \\
           +\,C\bigg(\!\usegraph{9}{delta322c}\bigg) &
              \Delta\bigg(\!\usegraph{9}{delta322i}\bigg)\,
      \Bigg] ,
\label{AllPlus4point2loopReduced}
\eeal
where the colour factor of the double-box irreducible numerator~\eqref{delta331}
is now explicit and shared with the butterfly contribution.
Apart from being more concise, interestingly, this representation coincides with the one obtained in the original calculation~\cite{Bern:2000dn}.

%%%%%%%%%%%%%%%%%%%%%%%%%%%%%%%%%%%%%%%%%%%%%%%%%%%%
\section{DDM-based colour decomposition}
\label{sec:DDMbased}
%%%%%%%%%%%%%%%%%%%%%%%%%%%%%%%%%%%%%%%%%%%%%%%%%%%%

The lesson that we can learn from the example in \sec{sec:4point2loop}
is that the trace-based colour decomposition~\eqref{Atraces}
runs over an overcomplete set of ordered cut topologies,
in the same way as the trace decomposition
of the tree amplitude~\eqref{AtreeTraces} involves amplitudes
that are linearly dependent under KK relations~\cite{Kleiss:1988ne}:
\be
   A(1,\alpha,n,\beta) = (-1)^{|\beta|} \!\!
      \sum_{\sigma \in\:\!\alpha \shuffle \beta^T} \!\!
      A(1,\sigma,n) \,.
\label{KK}
\ee
These relations resolve any colour-ordered amplitudes in terms of the subset of
$(n-2)!$ amplitudes with two particles fixed in colour-adjacent positions.
If one plugs this solution back to into
the trace decomposition~\eqref{AtreeTraces},
the resulting colour coefficients turn out to combine into
single strings of structure constants:
\beal
   {\cal A}^{(0)}_n = g^{n-2} \!\sum_{\sigma \in S_{n-2}}\!
      \tf^{\,a_1 a_{\sigma(2)} b_1} \tf^{\,b_1 a_{\sigma(3)} b_2} \dots
      \tf^{\,b_{n-4} a_{\sigma(n-2)} b_{n-3}}
      \tf^{\,b_{n-3} a_{\sigma(n-1)} a_n} & \\ \times
      A(1,\sigma(2),\dots,\sigma(n-1),n) & \\
    = g^{n-2} \!\sum_{\sigma \in S_{n-2}}\!
      C\bigg(\,\usegraph{10}{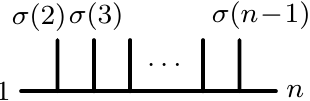}\,\bigg)\,
      A(1,\sigma(2),\dots,\sigma(n-1),n) &.
\label{AtreeDDM}
\eeal
This is the DDM colour decomposition~\cite{DelDuca:1999ha,DelDuca:1999rs},
and with respect to \eqn{AtreeTraces}
it avoids $(n-2)!(n-3)/2$ colour-ordered amplitudes.
Its four-point version, \eqn{AtreeDDM4point},
when applied inside the two-loop butterfly numerator~\eqref{delta330DDM},
relates the two decompositions of \sec{sec:4point2loop}.

%%%%%%%%% Figure %%%%%%%%%%%%%%%
\begin{figure}[t]
\centering
\includegraphics[scale=1.0]{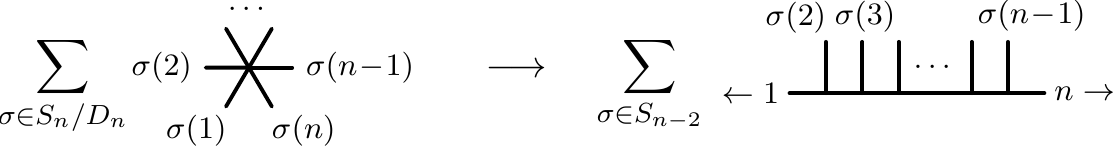}
\caption{\small Graphic representation of the process
of obtaining the DDM colour factors from traces via
``stretching'' of the vertex by any two edges (here chosen to be $1$ and $n$).}
\label{fig:TreeColourFactors}
\end{figure}
%%%%%%%%%%%%%%%%%%%%%%%%%%%%%%%%

More generally,
once the KK relations for irreducible numerators are taken into account
and their coefficients combined into the DDM-based colour factors,
we can rewrite the trace-based decomposition~\eqref{Atraces} as
\be
   {\cal A}^{(L)}_n = i^{L-1} g^{n+2L-2}\!\!\!\!\!\!\!
      \sum_{i\,\in\,\text{KK-independent\,1PI\,graphs}}
      \int\!\frac{\d^{LD} \ell}{(2\pi)^{LD}}
      \frac{C_i\,\Delta_i}{S_{\cal I} \prod_{l \in i} D_l} ,
\label{Areduced}
\ee
where we have also reabsorbed factors of $i$ into the numerators to produce
the overall factor of $i^{L-1}$.
The colour factors $C_i$ will be naturally given by trivalent diagrams,
but their concrete forms depend on the way the KK reduction is realised,
\ie on the choice of the KK-independent subset of ordered 1PI graphs.
The process of elimination the redundant topologies involves
picking two edges of a higher-point vertex and ``stretching'' it
by these edges to produce a sum over DDM ``half-ladder'' colour factors,
as illustrated by \fig{fig:TreeColourFactors},
each multiplied by its corresponding permutation of the original vertex.
The resulting subset of ordered topologies thus depends
on the choice of the ``stretched'' edges.

In the following sections, we shall demonstrate how to achieve this
automatically by inserting the DDM tree decomposition rather than the
trace basis decomposition.
But first, let us generalise the reasoning of \sec{sec:4point2loop}
on the validity of the KK relations for irreducible numerators.

%%%%%%%%%%%%%%%%%%%%%%%%%%%%%%%%%%%%%%%%%%%%%%%%%%%%
\subsection{Kleiss-Kuijf relations for irreducible numerators}
\label{sec:KK}
%%%%%%%%%%%%%%%%%%%%%%%%%%%%%%%%%%%%%%%%%%%%%%%%%%%%

In this section we argue that irreducible numerators can be chosen so
as to inherit the same KK relations from the colour-ordered unitarity
cuts, from which they are derived.

In the top-down approach, the definition of an irreducible numerator
involves subtraction of the higher-level numerators from its cut,
as summarised in \eqn{LevelSubtractionOrdered}.
Let us rewrite it with explicit reference to the level in the graph hierarchy,
labelled by $l,l+1,\dots,l_\text{max}$:
\be
   \Delta_i^{(l)}=
      \text{Cut}_i^{(l)} - \!\!
      \sum_{\text{uncut}\,D_{i\:\!j}} \!\!
      \frac{\Delta_{i\:\!j}^{(l+1)}}{D_{i\:\!j}} - ~\dots~ - \!\!
      \sum_{\text{uncut}\,D_{i\:\!j} \dots D_{i\:\!k}} \!\!
      \frac{\Delta_{i\:\!j \dots k}^{(l_\text{max})}}{D_{i\:\!j}
                                                \dots D_{i\:\!k}} .
\label{DeltasInCut}
\ee
Here every $\Delta_{i\:\!j \dots k}^{(l')}$ corresponds
to a cut obtained from $\text{Cut}_i^{(l)}$
by constraining additional loop propagators $D_{i\:\!j},\dots,D_{i\:\!k}$;
it is subtracted from $\text{Cut}_i^{(l)}$
to ensure polynomial nature of $\Delta_i^{(l)}$.
Obviously, the top-level numerators $\Delta_i^{(l_\text{max})}$
obey the KK relations of $\text{Cut}_i^{(l_\text{max})}$
on their cut kinematics,
as they correspond to maximal cuts without any subtractions.
In order for this to be true for general (off-shell) kinematics, we
choose the same basis of numerator functions for each term in a KK
relation, effectively associating the basis with the cut condition. In
this way, the KK relation of the irreducible numerators follows from
that of the cuts. Having established the
validity of the KK relations for top-level numerators, we shall now
construct a recursive argument for the lower levels.

Our argument will rely heavily on
the observation~\cite{DelDuca:1999rs,Bern:2008qj} that the KK relations
can be viewed as a consequence of the antisymmetry of a cubic expansion
for colour-ordered amplitudes.
Indeed, since colour content of a gauge theory consists
only of trivalent structures $\tf^{abc}$ and $T^a_{i\bar{\jmath}}$,
any Feynman diagram with insertions of higher-point Feynman vertices
can be absorbed into purely cubic diagrams,
modifying their kinematic (reducible) numerators $n_i$
but leaving any gauge-invariant object unchanged.
This ensures the existence of a cubic expansion for any tree or loop amplitude.
The antisymmetry under a trivalent-vertex flip
\be
   C_i = -C_j ~~~~~\Leftrightarrow~~~~~ n_i = -n_j
\label{fflip}
\ee
in this context simply means that $i$ and $j$ are essentially one cubic graph.
At tree level this means that, after reduction to colour traces,
different colour-ordered amplitudes pick up the same cubic graph
with different signs. The KK relations,
being purely colour-algebra statements without any reference to kinematics,
hold simply due to cancellation among the kinematic parts of the same cubic graphs occurring twice with opposite signs~\cite{DelDuca:1999rs,Bern:2008qj}.

Now consider a linear combination of cuts that vanishes due to a KK relation:
\beal
   0 = \sum_{\sigma_j} A(\sigma_j) \propto \sum_{i} \text{Cut}_i^{(l)} =
       \sum_{i} \Delta_i^{(l)} ~+~
     & \sum_{i} \sum_{\text{uncut}\,D_{i\:\!j}~~~\;}
       \!\!\!\!\!\!\!\!\!
       \frac{\Delta_{i\:\!j}^{(l+1)}}{D_{i\:\!j}} ~+~ \dots \\ ~+~
     & \sum_{i} \sum_{\text{uncut}\,D_{i\:\!j} \dots D_{i\:\!k}\!\!\!\!\!}
     ~ \frac{\Delta_{i\:\!j \dots k}^{(l_\text{max})}}{D_{i\:\!j}
                                                 \dots D_{i\:\!k}} .
\label{KKcut}
\eeal
This means that the cuts $\text{Cut}_i^{(l)}$
contain a common, up to permutations, vertex $j$
such that the corresponding colour-ordered tree amplitudes $A(\sigma_j)$ produce
pairwise annihilating cubic diagrams.
Supposing that all the KK relations for the higher-level numerators
are already proven, we can assume that for all such irreducible numerators
$\Delta_{i\:\!j \dots k}^{(l'>l)}$ on the right-hand side
there exist an expansion in purely cubic graphs.
We will not need any specific form of such kinematic diagrams;
all we are going to require is their existence.
These cubic graphs must contain subgraphs that correspond
to permutations of the same vertex $j$. Indeed,
the subtraction procedure that generated those higher levels in the first place
started with the cut graphs containing that vertex and then proceeded
by exposing loop-dependent propagators inside of it and other vertices.
Therefore, when each of the generated higher-level graphs
is expanded into purely cubic graphs, they will all contain
the propagators of the original cut
and hence the subgraphs corresponding to the vertex $j$.
From now on, we can concentrate on what is happening inside these subgraphs.

Consider the cubic subgraphs generated by two numerators
of adjacent levels $l'$ and $l'+1$.
Since the latter's graph contains one more exposed propagator
than the former's from the start,
this propagator will be present in all of its cubic graphs,
thus picking a subset of the cubic graphs of the former.
Hence each cubic subgraph produced by any higher-level numerator
is isomorphic to some cubic subgraph of the vertex $j$.
But the permutation sum on the left-hand side guaranteed
that every cubic subgraph will appear in expansion of
two (or an even number of) colour-ordered amplitudes with opposite signs.
Since the subtraction procedure was only sensitive to the propagator structure,
any cubic subgraph produced by some higher-level numerator
must be accompanied by a twin with the opposite orientation,
produced by another higher-level numerator.
In other words, the fact that the higher-level numerators
$\Delta_{i\:\!j \dots k}^{(l'>l)}$ do not contain all possible cubic graphs
cannot interfere with the cancellations coming from the overall permutation sum.
Therefore, we conclude that the cubic diagrams must cancel at all higher levels, and then the equality~\eqref{KKcut} itself proves
the KK relation for the level at hand, as needed for our inductive argument.

More formally,
if $ {\cal C} \equiv \{\Gamma_j \in \sum_{i} \text{Cut}_i^{(l)} \} $
is the set of all oriented cubic graphs on the left-hand side of \eqn{KKcut},
then it can be divided into two equal sets
$ {\cal C}^+ \cup {\cal C^-} =  {\cal C} $ that differ only by orientation.
For each level $l'>l$ on the right-hand side,
there is a map of the numerator graphs $\Gamma_j^{(l')}$
to a subset of ${\cal C}$, representing the trivalent expansion
of the higher-level numerators.
By construction, since it only involves exposing loop-dependent propagators,
this map is symmetric with respect to both orientations
${\cal C}^+$ and ${\cal C^-}$.
Therefore, for
$ \forall\,\Gamma_j^{(l')} \in \sum_{i;j,\dots,k}
  \Delta_{i\:\!j \dots k}^{(l')}/(D_{i\:\!j} \dots D_{i\:\!k}) $,
there $ \exists (- \Gamma_j^{(l')}) \in \sum_{i;j,\dots,k}
  \Delta_{i\:\!j \dots k}^{(l')}/(D_{i\:\!j} \dots D_{i\:\!k}) $
with an opposite orientation.
This guarantees cancellation of the cubic diagrams from all levels, except $l$,
and the equality~\eqref{KKcut} itself then proves the KK relation
for $\Delta_i^{(l)}$, closing the induction.

%%%%%%%%%%%%%%%%%%%%%%%%%%%%%%%%%%%%%%%%%%%%%%%%%%%%
\subsection{One-loop case}
\label{sec:1loop}
%%%%%%%%%%%%%%%%%%%%%%%%%%%%%%%%%%%%%%%%%%%%%%%%%%%%

In this section we consider the application
of the DDM-based colour decomposition~\eqref{Areduced}
to a purely gluonic one-loop amplitude.
As one would expect,
we recover the one-loop DDM colour decomposition~\cite{DelDuca:1999rs}:
\beal
   {\cal A}^{(1)}_n
    = g^{n} \!\!
      \sum_{\sigma\in S_n/D_n} \!\!
      \tf^{\,b_1 a_{\sigma(1)} b_2} \tf^{\,b_2 a_{\sigma(2)} b_3} \dots
      \tf^{\,b_n a_{\sigma(n)} b_1}\,
      A^{(1)}(\sigma(1),\sigma(2),\dots,\sigma(n)) & \\
    = g^{n} \!\!
      \sum_{\sigma\in S_n/D_n} \!\!
      C\Bigg(\usegraph{20}{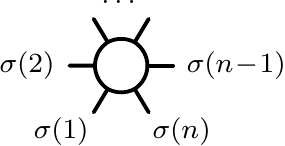}\Bigg)\,
      A^{(1)}(\sigma(1),\sigma(2),\dots,\sigma(n)) & ,
\label{A1loopDDM}
\eeal
where $A^{(1)}$ are leading-colour partial amplitudes.
Recall that these objects can be defined at any loop order as gauge-invariant coefficients of the leading single-trace colour factors:
\beal \!\!\!\!\!
   {\cal A}^{(L)}_n = i^{L-1} g^{n+2L-2} \Big\{ N_c^L \!\!
      \sum_{\sigma\in S_n/Z_n} \!\!
      \Tr\!\big(T^{a_{\sigma(1)}} T^{a_{\sigma(2)}} \dots T^{a_{\sigma(n)}}\big)
      A^{(L)}(\sigma(1),\sigma(2),\dots,\sigma(n)) & \\
    +~{\cal O}\big(N_c^{L-1}\big) & \Big\} . \!\!
\label{leadingcolour}
\eeal
At one loop, all the subleading partial amplitudes are known to be linear combinations of the leading ones~\cite{Bern:1990ux},
and thus the full amplitude can be written in terms of $A^{(1)}$ only,
as implemented by the one-loop DDM decomposition~\eqref{A1loopDDM}.

\begin{figure}[t]
   \centering
   \includegraphics[width=0.73\textwidth]{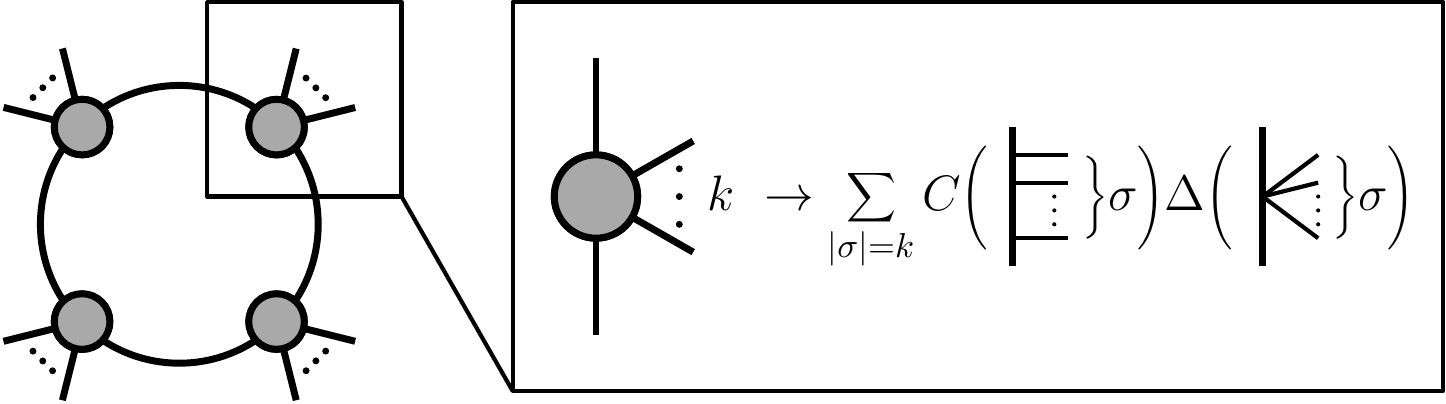}
   \caption{\small Inserting the DDM tree basis into coloured cuts
of a one-loop amplitude. As shown in the inset, ``stretching'' by the two internal edges gives a sum over permutations of the external legs.}
\label{fig:color1loop1topo}
\end{figure}

To see how \eqn{A1loopDDM} follows from \eqn{Areduced}, let us
consider a generic one-loop topology depicted in \fig{fig:color1loop1topo}.
In $D=4-2\epsilon$ dimensions, it consists of up to five vertices connected into a loop,
each with 3 or more edges, two of which are internal.
To use our DDM-based prescription, it is natural to use these internal edges
as the ``stretching'' ends for the DDM colour factors,
so that the resulting ordered graphs are enumerated by permutations
of external legs on one side of the resulting ``half-ladder''.
This procedure eliminates ordered topologies
with some of the external legs sticking inside the loop;
they correspond to valid colour-ordered cuts
but are redundant under the KK relations.
Therefore, the remaining ordered topologies are
pentagons, boxes, triangles, bubbles and tadpoles
with all external particles looking, for definiteness, outside the loop.
Moreover, all colour factors will be loops of structure constants,
as in \eqn{A1loopDDM}, differing only by permutations of external legs.
The ordered topologies that will contribute to the kinematic coefficient of
$\tf^{\,b_1 a_{1} b_2} \tf^{\,b_2 a_{2} b_3} \dots \tf^{\,b_n a_{n} b_1}$,
for instance, will be precisely the colour-ordered amplitude\footnote{For generality,
we allowed tadpoles and bubbles on external legs
in the one-loop adjoint-representation amplitude~\eqref{leadingcolour1loop},
even though in dimensional regularisation they integrate to zero
in the massless case.}
\beal \!\!\!\!\!
   A^{(1)}(1,2,\dots,n) = I
      \Bigg[ \sum_{\substack{1\le i_1<i_2<i_3\quad\\
                             \quad~\;\:<i_4<i_5\le n}}\!\!\!\!
             \Delta\!\left(\!\scalegraph{0.33}{70}{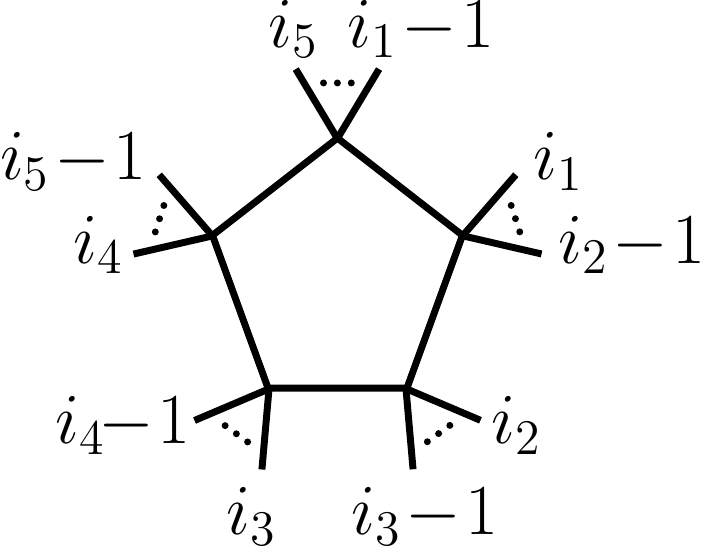}\!\!\right)
    +\!\!\!\!\sum_{\substack{1\le i_1<i_2~~\\
                             ~~\,<i_3<i_4\le n}}\!\!\!\!
             \Delta\!\left(\!\scalegraph{0.33}{64}{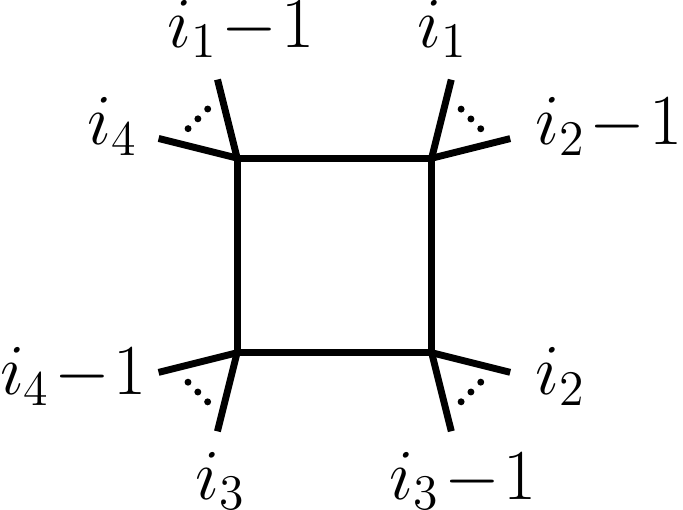}\!\right) \\
          +\!\sum_{1\le i_1<i_2<i_3\le n}\!\!\!\!
             \Delta\!\left(\!\scalegraph{0.33}{55}{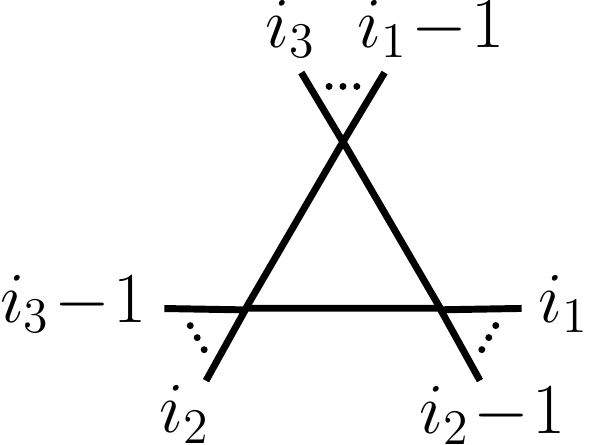}\!\right)
          +\!\sum_{1\le i_1<i_2\le n}\!\!\!
             \Delta\Big(\scalegraph{0.33}{9}{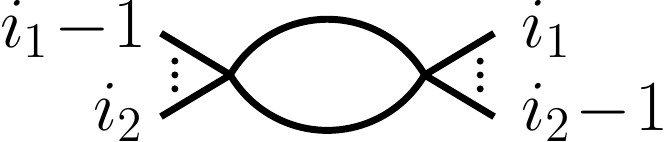}\Big)
          +\!\sum_{1\le i_1\le n}\!\!
             \Delta\Big(\scalegraph{0.33}{9}{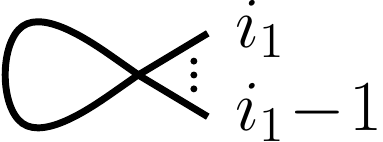}\Big) &
      \Bigg] . \!\!\!\!\!
\label{leadingcolour1loop}
\eeal
The only subtlety here is that fixing the two internal edges
inside the bubble topologies results in pairs of equivalent ordered topologies.
For example, the two DDM decompositions inside the coloured bubble $(1,2,3;4,5)$ produce the orderings $(1,2,3;4,5)$ and $(3,2,1;5,4)$ as distinct permutations,
though obviously with equivalent colour factors
$\pm \tf^{\,b_1 a_{1} b_2} \tf^{\,b_2 a_{2} b_3} \tf^{\,b_3 a_{3} b_4}
                           \tf^{\,b_4 a_{4} b_5} \tf^{\,b_5 a_{5} b_1}$.
This precisely cancels the bubble symmetry factor $S_\text{bub}=2$,
thereby absent from the leading-colour amplitude~\eqref{leadingcolour1loop}.
A similar argument takes care of the tadpole symmetry factor $S_\text{tad}=2$.
This concludes the check that at one loop
our DDM-induced multi-loop decomposition~\eqref{Areduced} reduces to
the proper DDM decomposition~\eqref{A1loopDDM}.

%%%%%%%%%%%%%%%%%%%%%%%%%%%%%%%%%%%%%%%%%%%%%%%%%%%%
\subsection{Two-loop case}
\label{sec:2loop}
%%%%%%%%%%%%%%%%%%%%%%%%%%%%%%%%%%%%%%%%%%%%%%%%%%%%

\begin{figure}[t]
   \centering
   \includegraphics[width=0.91\textwidth]{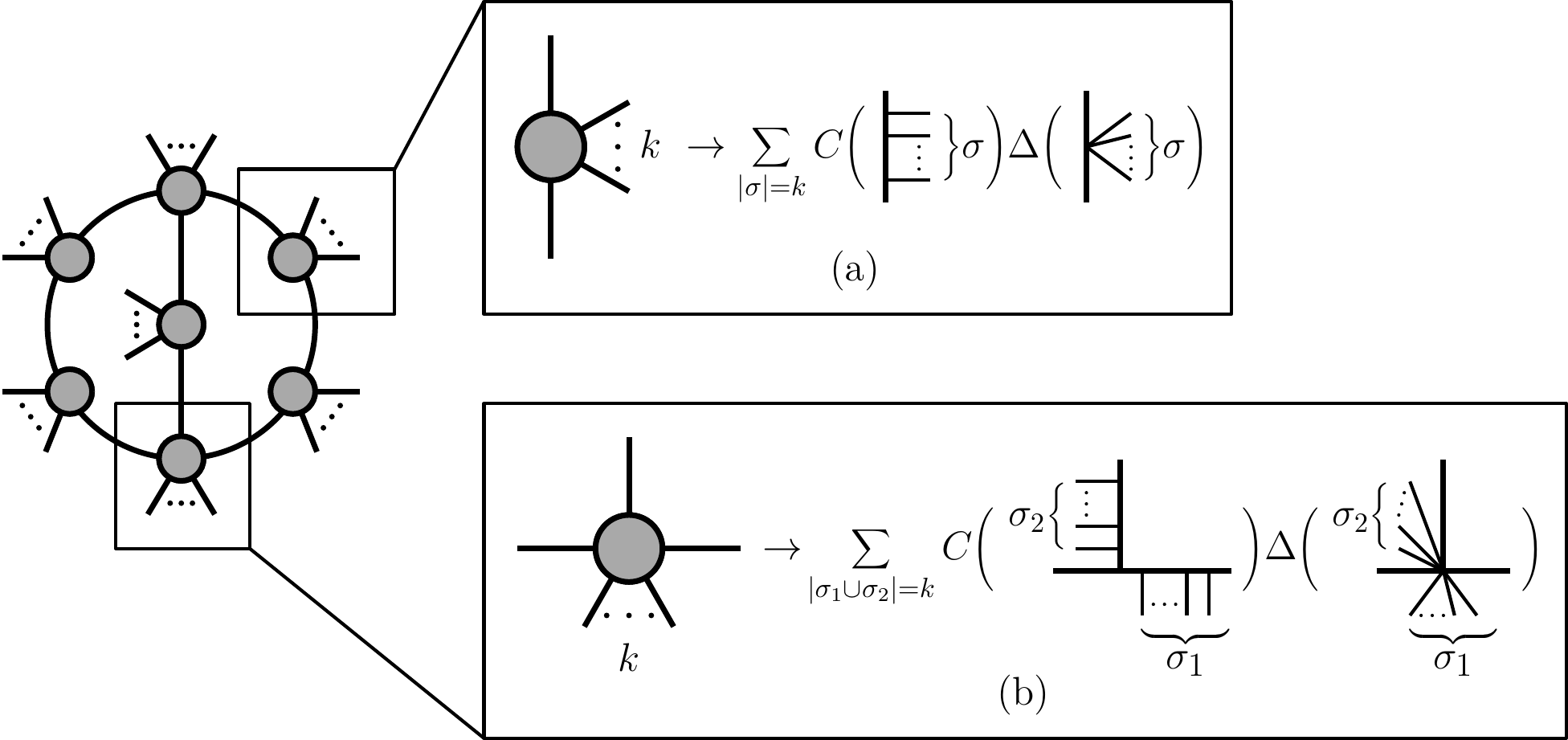}
   \caption{\small Inserting the DDM tree basis into a pure two-loop coloured cut. The upper inset~(a) shows the case of two loop-dependent legs (thus equivalent to the one-loop decomposition in \fig{fig:color1loop1topo}), while the lower inset~(b) shows the case of three loop-dependent legs. The sums run over the permutations of the external legs in the tree-level amplitude.}
   \label{fig:color2looppuretopo}
\end{figure}

Let us now consider the case of the two-loop colour decomposition
by cuts~\cite{Badger:2015lda}.
At two loops, there are two basic topologies:
the ``pure two-loop'' topology and the ``butterfly'',
given in \Figs{fig:color2looppuretopo}{fig:color2loopbutterflytopo}, respectively.
In order to find a KK-independent basis of colour-ordered objects,
we wish once more to insert the DDM decomposition into the coloured cuts.
As at one loop we must make choices of which two gluons to hold fixed
in the DDM tree.
Here the procedure is more complicated
as some coloured trees contain more than two loop-dependent legs.
We find trees with two and three loop-dependent legs in the pure two-loop topologies
and two and four loop-dependent legs in the butterfly topologies.
Let us consider these cases in turn.

\begin{figure}[t]
   \centering
   \includegraphics[width=0.91\textwidth]{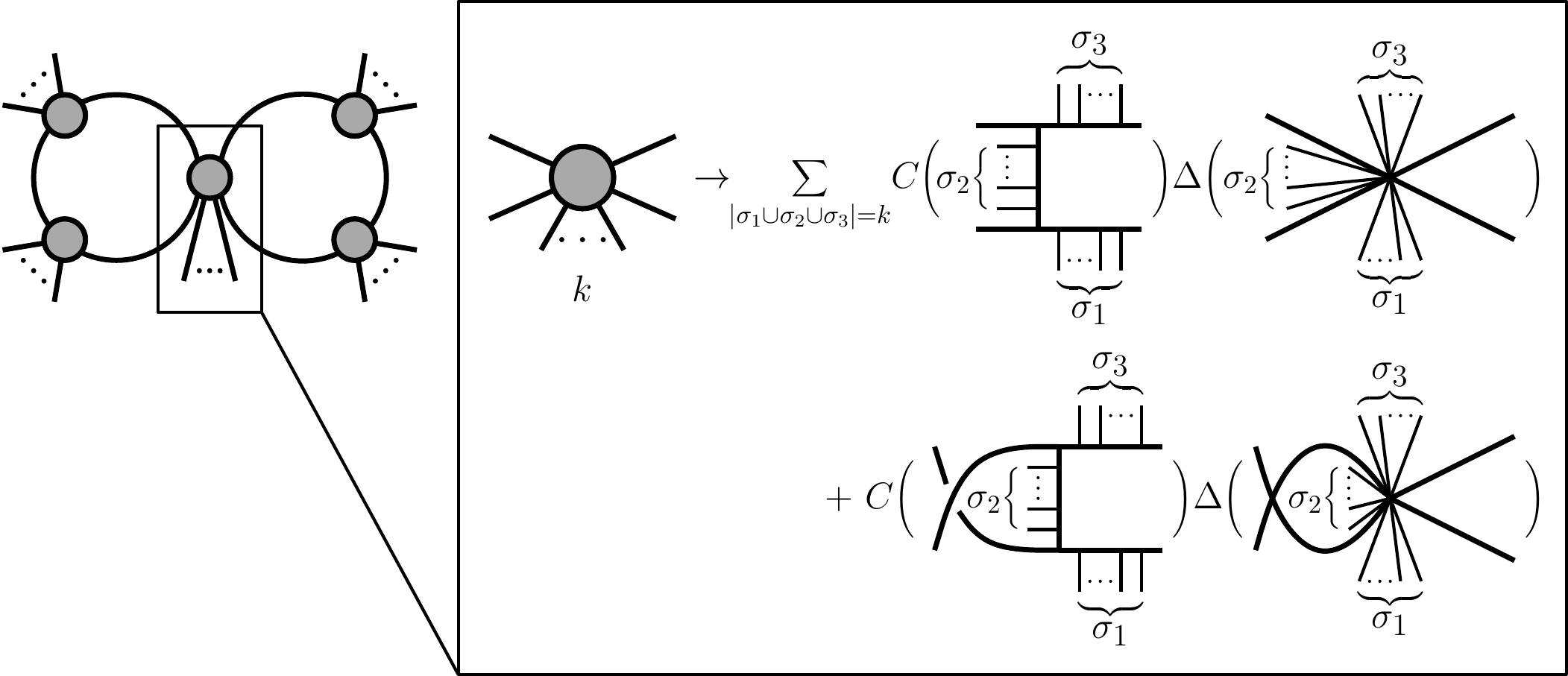}
   \caption{\small Inserting the DDM tree basis into a coloured cut of the butterfly topology. There are four loop edges, and the inset shows the result of inserting the DDM tree decomposition fixing the right ones. The sums run over the permutations of the external legs in the tree-level amplitude.}
   \label{fig:color2loopbutterflytopo}
\end{figure}

\begin{itemize}
\item Two loop legs:
As at one loop, perform the DDM decomposition
holding the two loop edges fixed and permuting over the external legs.
See inset~(a) of \fig{fig:color2looppuretopo}.
\item Three loop legs:
Choose two loop edges to fix and DDM-decompose,
permuting across the external legs and the third loop edge.
See inset~(b) of \fig{fig:color2looppuretopo}.
\item Four loop legs:
Choose one side on which to fix two of the loop edges and
permute over the external legs and the two loop edges on the other side.
See \fig{fig:color2loopbutterflytopo}.
\end{itemize}

The canonical choices made at two loops remove some of the symmetry properties of the coloured diagrams in order to ensure KK independence.
Note that there is a degree of arbitrarity in this procedure,
which was not present at one loop,
where every tree in a cut has exactly two loop-dependent legs.

It should also be noted that a KK-independent numerator basis at two loops does not enjoy the complete consistency between levels which is found at one loop.
Specifically, when performing the hierarchy subtraction of a lower-level topology, one may run into an ancestor graph that was not in the KK-independent basis defined by the DDM-based colour decomposition of the coloured ancestor numerator.
However, as the terms in that decomposition form a KK-independent basis, one can always use the KK relations to obtain the colour-ordered ancestor, needed for lower-level subtraction.
For example, if the canonical procedure in \fig{fig:color2looppuretopo}b
colour-decomposes the following double-triangle topology as
\beal \!\!\!\!\!
   \tilde{\Delta}\!\left(\!\scalegraph{0.37}{37}{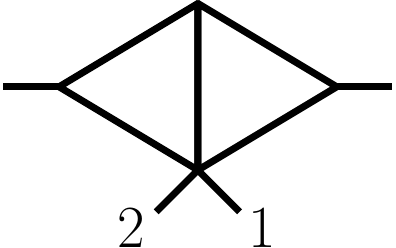}\!\right)
 = C\!\left(\!\scalegraph{0.37}{31}{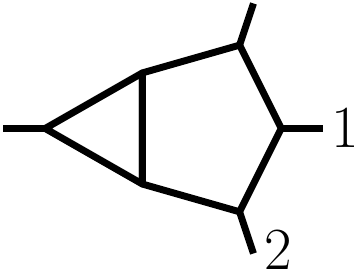}\!\! \right)
   \Delta\!\left(\!\scalegraph{0.37}{37}{delta221i}\!\right)
 + C\!\left(\!\scalegraph{0.37}{31}{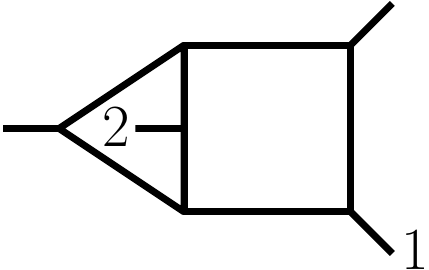}\:\!\right)
   \Delta\!\left(\!\scalegraph{0.37}{46}{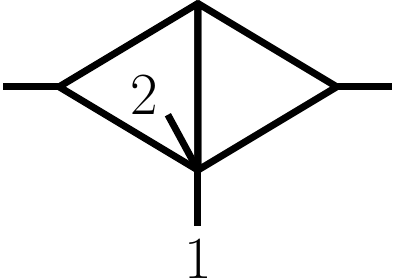}\!\right) & \\
+\,C\!\left(\!\scalegraph{0.37}{26}{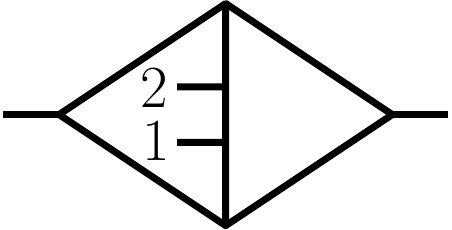}\!\right)
   \Delta\!\left(\!\scalegraph{0.37}{18}{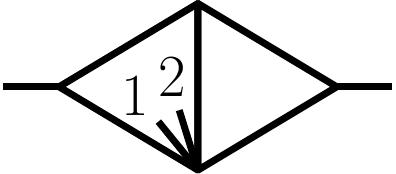}\!\right) &
 + \{1\leftrightarrow2\} , \!\!\!\!\!
\label{eq:semisimpletriangletriangleexpansion}
\eeal
where we see the colour-ordered numerator
$\Delta\big(\scalegraph{0.67}{10}{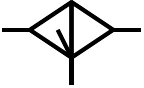}\big)$.
Its hierarchy subtraction naturally involves the numerator
$\Delta\big(\scalegraph{0.67}{4.5}{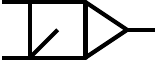}\big)$,
which is excluded by the usual procedure in \fig{fig:color2looppuretopo}a
in favour of its planar-looking orderings, with the underlying KK relation being
\beal
   \Delta\!\bigg(\;\!\scalegraph{0.37}{39}{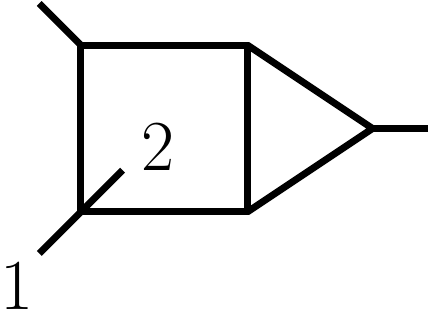}\:\!\!\bigg)
 =-\Delta\!\left(\scalegraph{0.37}{48}{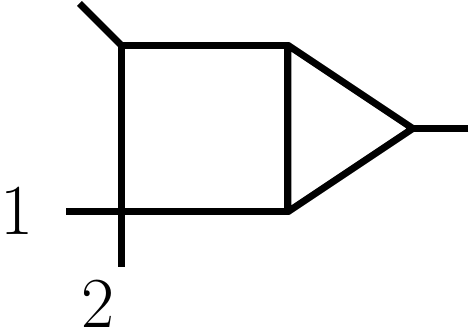}\!\right)
 - \Delta\!\left(\scalegraph{0.37}{48}{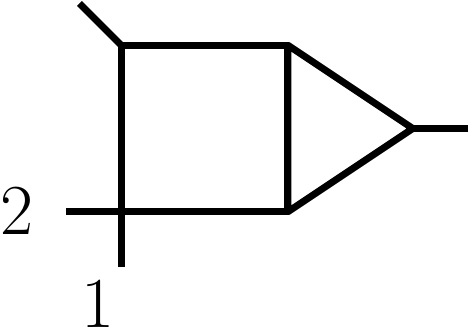}\!\right) .
\label{eq:semisimpletriangletriangleparent}
\eeal
Therefore, one can perform the hierarchy subtraction for
$\Delta\big(\scalegraph{0.67}{10}{delta221NPc1}\big)$
plugging in \eqn{eq:semisimpletriangletriangleparent}
whenever $\Delta\big(\scalegraph{0.67}{4.5}{delta321NPc1}\big)$ is needed.
To sum up, one must make sure to always KK-reduce the ancestor expressions to the basis of previously computed numerators $\Delta_i$.

%%%%%%%%%%%%%%%%%%%%%%%%%%%%%%%%%%%%%%%%%%%%%%%%%%%%
\subsubsection{Two-loop symmetry factors}
\label{sec:symmetry2loop}
%%%%%%%%%%%%%%%%%%%%%%%%%%%%%%%%%%%%%%%%%%%%%%%%%%%%

For coloured cuts with non-unity symmetry factors,
one can often choose a KK-independent basis of colour-ordered cuts
which is redundant under loop-momentum relabelling.
Therefore, when this colour decomposition is projected
onto irreducible numerators, which are momentum-routing-invariant under the integral sign,
the basis is further reduced.
We shall now demonstrate at two loops that in all cases where this is possible, the degree of the redundancy is equivalent to the symmetry factor,
and many symmetry factors can be cancelled in this approach.
However, the canonical choices
in \figs{fig:color2looppuretopo}{fig:color2loopbutterflytopo}
do not always achieve this,
in contrast to the one-loop case,
where the usual stretch choice (shown in \fig{fig:color1loop1topo})
is enough to find a simple momentum-routing redundant basis.

Consider the most symmetric case of the sunrise diagram
$\Delta\big(\scalegraph{0.37}{10}{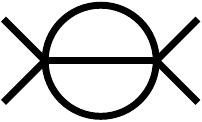}\big)$,
which has a $3!$ permutation symmetry (or $(L+1)!$ in the $L$-loop case
of $\Delta\big(\scalegraph{0.37}{10}{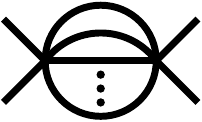}\big)$). Evidently,
the canonical DDM stretching procedure would single out two loop legs,
so the full propagator-permutation symmetry would not be accounted for.
This is why in this case it is more useful to
stretch across the two \emph{external} legs on both trees.
This results in $3!\cdot 3! = 36$ different terms,
all of them being different permutations of the loop lines.
However, there are only 6 distinct topologies,
each appearing 6 times due to $3!$ different labellings of the propagators.
As $\Delta_i$ are label-invariant, these terms can be collected,
leading to an overall factor of 6, cancelling the symmetry factor.
Therefore, this procedure gives
\beal
   \frac{1}{6}
   \tilde{\Delta}\!\left(\:\!\!\scalegraph{0.4}{10}{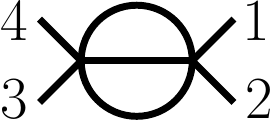}\:\!\!\right)
 = C\!\left(\scalegraph{0.37}{25}{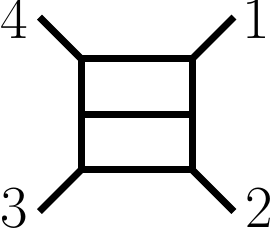}\right)
   \Delta\!\left(\:\!\!\scalegraph{0.4}{10}{sunrise1234}\:\!\!\right)
 + C\!\left(\scalegraph{0.37}{25}{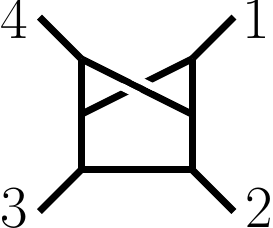}\right)
   \Delta\!\left(\:\!\!\scalegraph{0.4}{10}{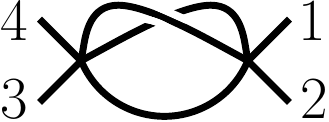}\:\!\!\right) & \\
+\,C\!\left(\scalegraph{0.37}{25}{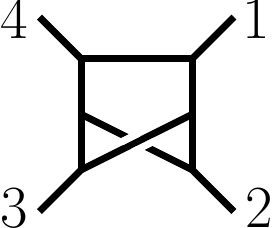}\right)
   \Delta\!\left(\:\!\!\scalegraph{0.4}{10}{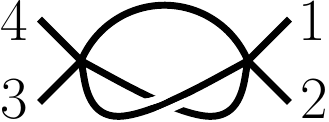}\:\!\!\right) &
 + \{3\leftrightarrow4\} .
%+\,C\!\left(\scalegraph{0.37}{25}{sunrise1243c}\right)
%   \Delta\!\left(\:\!\!\scalegraph{0.4}{10}{sunrise1243}\:\!\!\right)
% + C\!\left(\scalegraph{0.37}{25}{sunriseNPup1243c}\right)
%   \Delta\!\left(\:\!\!\scalegraph{0.4}{10}{sunriseNPup1243}\:\!\!\right)
% + C\!\left(\scalegraph{0.37}{25}{sunriseNPdown1243c}\right)
%   \Delta\!\left(\:\!\!\scalegraph{0.4}{10}{sunriseNPdown1243}\:\!\!\right) & ,
\label{eq:sunriseexpansion}
\eeal
This argument can be repeated for more than two external legs
on the sides of the sunrise-type diagrams,
in which case the permutation sum will still cancel the symmetry factor
but contain ordered configurations of the extra external legs
pointing inside the loops.

\begin{figure}
   \centering
\begin{tabular}{ c c c c c c c}
   \scalegraph{0.5}{4}{sunrise} &
   \scalegraph{0.5}{0}{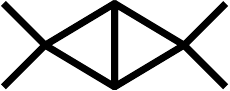} &
   \scalegraph{0.5}{0}{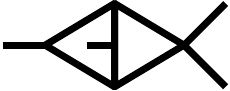} &
   \scalegraph{0.5}{0}{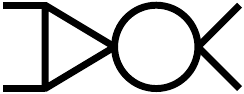} &
   \scalegraph{0.5}{0}{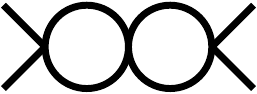} &
   \scalegraph{0.5}{5}{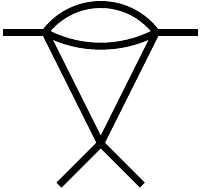} &
   \scalegraph{0.5}{5}{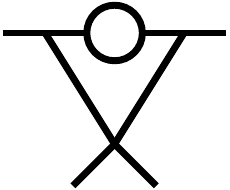} \\
   6 & 2 & 2 & 2 & 4 & 2 & 2\\
\end{tabular}
   \caption{Example 4-point coloured diagrams and their associated symmetry factors. In all but one case these factors can be made to cancel in the colour decomposition.}
   \label{fig:TwoLoopSymmetryFactors}
\end{figure}

At two loops, there are six more graphs with non-unity symmetry factors
listed in \fig{fig:TwoLoopSymmetryFactors}.
All other such graphs should be topologically similar to these.
For the planar and non-planar double triangle topologies,
the factor due to the overall reflection symmetry of the diagrams,
is cancelled by the same argument as for one-loop bubbles and tadpoles,
through the canonical loop-leg stretching.
In the case of the butterfly topologies, the canonical procedure
in \figs{fig:color2looppuretopo}{fig:color2loopbutterflytopo}
results in a permutation sum over the internal loop edges,
involving terms identical up to internal-edge relabelling.
Once again this cancels the symmetry factors,
leaving only two nonequivalent external leg configurations
with double-box colour factors in both cases.

To cancel the symmetry factor of the triangle-bubble,
one may canonically DDM-stretch the four-point vertices
holding the legs of the bubble fixed,
\beal \!\!\!
   \frac{1}{2}
   \tilde{\Delta}\!\left(\!\scalegraph{0.4}{22}{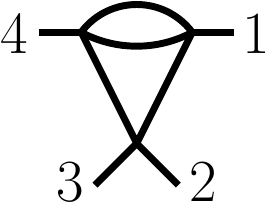}\!\right)
 = \frac{1}{2}
   \bigg\{
   C\!\left(\scalegraph{0.37}{25}{sunrise1234c}\right)
   \Delta\!\left(\!\scalegraph{0.4}{22}{trianglebubbleP}\!\right)
 + C\!\left(\scalegraph{0.37}{25}{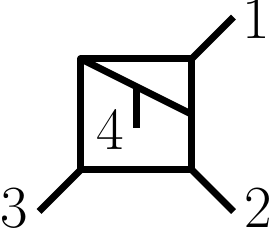}\right)
   \Delta\!\left(\scalegraph{0.4}{22}{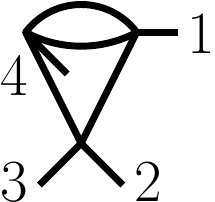}\!\right) \\
+\,C\!\left(\scalegraph{0.37}{25}{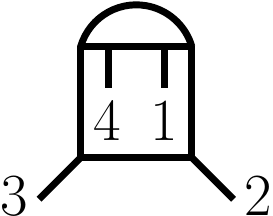}\right)
   \Delta\!\left(\scalegraph{0.4}{22}{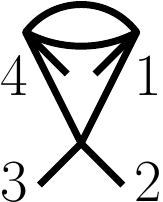}\right)
 + C\!\left(\scalegraph{0.37}{25}{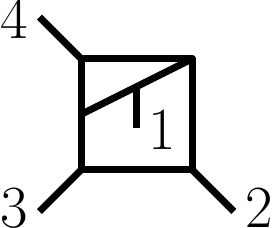}\right)
   \Delta\!\left(\!\scalegraph{0.4}{22}{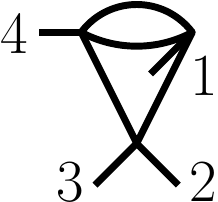}\right) &
 + \{2\leftrightarrow3\}
   \bigg\} , \!\!\! %\\
% = C\!\left(\scalegraph{0.37}{25}{sunrise1234c}\right)
%   \Delta\!\left(\!\scalegraph{0.4}{22}{trianglebubbleP}\!\right)
% + C\!\left(\scalegraph{0.37}{25}{trianglebubbleNPleftc}\right)
%   \Delta\!\left(\scalegraph{0.4}{22}{trianglebubbleNPleft}\!\right)
% + \,\,\,&\!\!\! \{2\leftrightarrow3\} .
\label{eq:trianglebubbleexpansion}
\eeal
with the vertex with legs $2$ and $3$ dealt with in the usual fashion.
Now the symmetry factor can be removed because the graphs in the second line
are equivalent to those shown in the first line due to
the reflection symmetry of the vertices.

Finally, we should note that there is a special case where
the symmetry factor cannot be cancelled
--- the diagram with the internal bubble associated with
propagator renormalisation. As the symmetric part of the diagram
is built out of three-point tree amplitudes,
there is only one term arising from the colour decomposition,
and so the symmetry factor remains.
This is consistent with the fact that in this particular type of diagram
the DDM-based colour factor respects the symmetry giving rise to that factor.

%%%%%%%%%%%%%%%%%%%%%%%%%%%%%%%%%%%%%%%%%%%%%%%%%%%%
\subsubsection{Five-point two-loop amplitude}
\label{sec:5point2loop}
%%%%%%%%%%%%%%%%%%%%%%%%%%%%%%%%%%%%%%%%%%%%%%%%%%%%

Let us consider an explicit two-loop example in order to further elucidate the technique. Here we show the decomposition of the first two levels of the five-point pure Yang-Mills amplitude for arbitrary helicity:
\small
\begin{align}
\label{5pointPureYMDecomposition}
   {\cal A}^{(2)}(1,2,3,4,5) =
   i g^7 & \sum_{\sigma \in S_5} \sigma \\ \circ\!\;
   I\Bigg[
      C\bigg(\eqnDelta{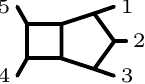}\!\!\bigg)
      \Bigg\{ &
      \frac{1}{2} \Delta\bigg(\eqnDelta{delta431i}\!\!\bigg)
    + \Delta\bigg(\;\!\usegraph{9.5}{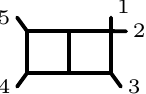}\!\:\bigg)
    + \frac{1}{2} \Delta\bigg(\!\;\usegraph{13.5}{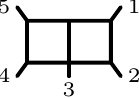}\!\;\bigg)
    + \Delta\bigg(\!\!\eqnDelta{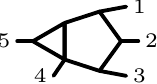}\!\!\bigg) \nn \\ & \qquad~\,
    + \frac{1}{2} \Delta\bigg(\!\!\!\;\eqnDelta{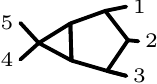}\!\!\bigg)
    + \frac{1}{2} \Delta\bigg(\eqnDelta{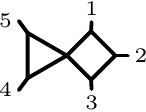}\!\!\bigg)
    + \frac{1}{2} \Delta\bigg(\eqnDelta{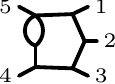}
                       +\!\!\!\eqnDelta{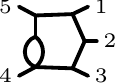}\!\!\!\:\bigg)
 \!   \Bigg\} \nn \\
   +\,C\bigg(\!\!\eqnDelta{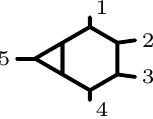}\!\!\!\;\bigg)
      \Bigg\{ &
      \frac{1}{2} \Delta\bigg(\!\!\eqnDelta{delta521i}\!\!\!\;\bigg)
    + \Delta\bigg(\!\!\!\;\usegraph{9}{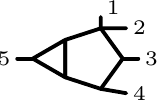}\!\!\!\;\bigg)
    + \frac{1}{2} \Delta\bigg(\!\!\eqnDelta{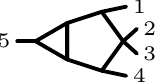}\!\!\!\:\bigg)
 \!   \Bigg\} \nn \\
   +\,\frac{1}{2} C\bigg(\eqnDelta{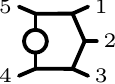}\!\!\!\:\bigg)
      \Bigg\{ &
      \frac{1}{2} \Delta\bigg(\eqnDelta{delta611i}\!\!\!\:\bigg)
    + \Delta\bigg(\!\:\usegraph{9.5}{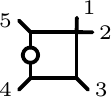}\bigg)
    + \Delta\bigg(\usegraph{9.5}{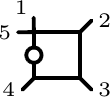}\!\:\bigg)
 \!   \Bigg\} \nn \\
   +\,C\bigg(\eqnDelta{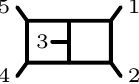}\bigg)
      \Bigg\{ &
      \frac{1}{4} \Delta\bigg(\eqnDelta{delta332NPi}\bigg)
    + \frac{1}{2} \Delta\bigg(\eqnDelta{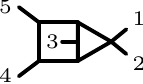}\!\bigg)
    + \frac{1}{2} \Delta\bigg(\eqnDelta{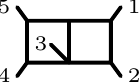}\bigg)
    + \Delta\bigg(\!\!\!\;\usegraph{16.5}{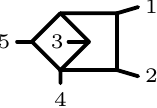}\!\:\bigg)
 \!   \Bigg\} \nn \\
   +\,C\bigg(\!\!\eqnDelta{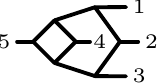}\!\!\bigg)
      \Bigg\{ &
      \frac{1}{4} \Delta\bigg(\!\!\eqnDelta{delta422NPi}\!\!\bigg)
    + \frac{1}{2} \Delta\bigg(\!\!\!\;\usegraph{8.8}{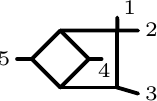}\!\:\bigg)
    + \Delta\bigg(\!\!\eqnDelta{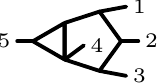}\!\!\bigg) \nn \\ &
      \qquad \qquad \qquad \qquad \qquad \quad~\:
    + \frac{1}{2} \Delta\bigg(\eqnDelta{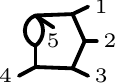}
                       +\!\!\!\eqnDelta{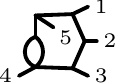}\!\!\!\:\bigg)
 \!   \Bigg\}
 \! + \dots \,
   \Bigg] . \nn
\end{align}
\normalsize
For the special case of all positive helicities, the corresponding irreducible numerators  can be found in \rcite{Badger:2013gxa}, where the colour decomposition goes one level lower, but omits some topologies at the first two levels that are shown in \eqn{5pointPureYMDecomposition}. Here we include all the topologies whose numerators do not have to vanish for arbitrary helicity.
Going to the next level in full generality is straightforward
but requires much more space than in the all-plus case~\cite{Badger:2013gxa}.

Note that in \eqn{5pointPureYMDecomposition} the rational prefactors
of the $\Delta$s are not symmetry factors, but factors that remove the
overcounting by summing over the elements of $S_5$. This is true even
for the diagrams that naively have symmetry factors (such as those
with internal bubbles) because these can be cancelled in the
decomposition, as explained in \sec{sec:symmetry2loop}. The only
genuine symmetry factor due to bubble insertion is shown in front of
the related colour factor.

A subtlety in the two-loop procedure comes from terms like
$\Delta\big(\,\scalegraph{0.67}{7}{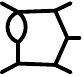}
           +\!\scalegraph{0.67}{7}{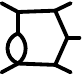}\big)$ and
$\Delta\big(\,\scalegraph{0.67}{7}{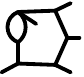}
           +\!\scalegraph{0.67}{7}{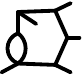}\big)$
--- the cuts associated to these topologies do not exist.
We have obtained their contribution
to the decomposition~\eqref{5pointPureYMDecomposition}
by taking the factorisation limit of the colour decomposition of
the lower-level topology
$\Delta\big(\!\eqnDiag{\scalegraph{0.64}{0}{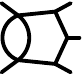}}\;\!\!\big)$
%(involving $\Delta\big(\!\eqnDiag{\scalegraph{0.64}{0}{delta411UNP2c}}\;\!\!\big)$)
which reveals the respective propagator,
see \app{app:subtlety} for more details.

%%%%%%%%%%%%%%%%%%%%%%%%%%%%%%%%%%%%%%%%%%%%%%%%%%%%
\subsection{Three-loop case}
\label{sec:3loop}
%%%%%%%%%%%%%%%%%%%%%%%%%%%%%%%%%%%%%%%%%%%%%%%%%%%%

%%%%%%%%% Figure %%%%%%%%%%%%%%%
\begin{figure}[t]
   \centering
   \includegraphics[width=\textwidth,trim=0 0 0 10pt]{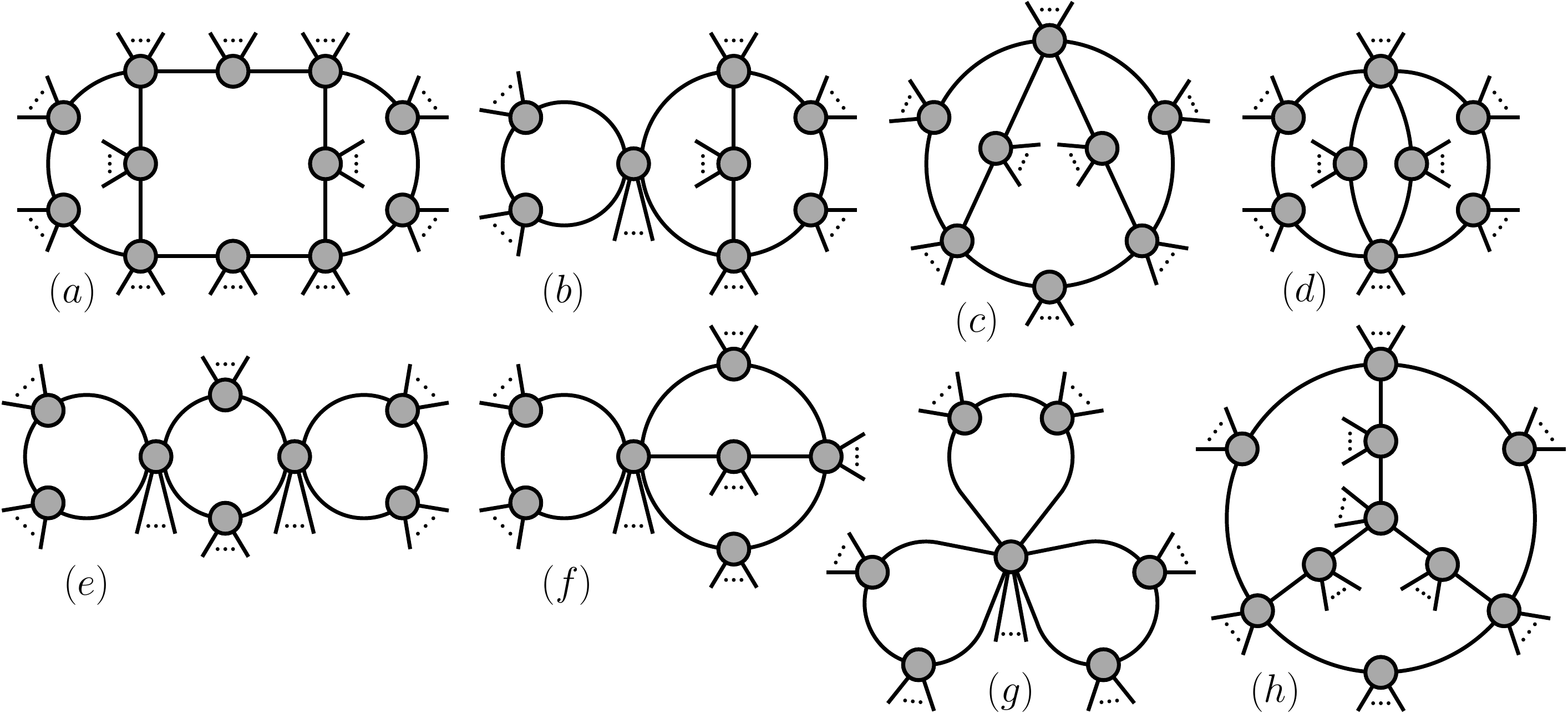}
   \caption{\small Basic topologies of three-loop coloured cuts.}
   \label{fig:topo3all}
\end{figure}
%%%%%%%%%%%%%%%%%%%%%%%%%%%%%%%%

\begin{figure}[t]
   \centering
   \includegraphics[width=0.82\textwidth]{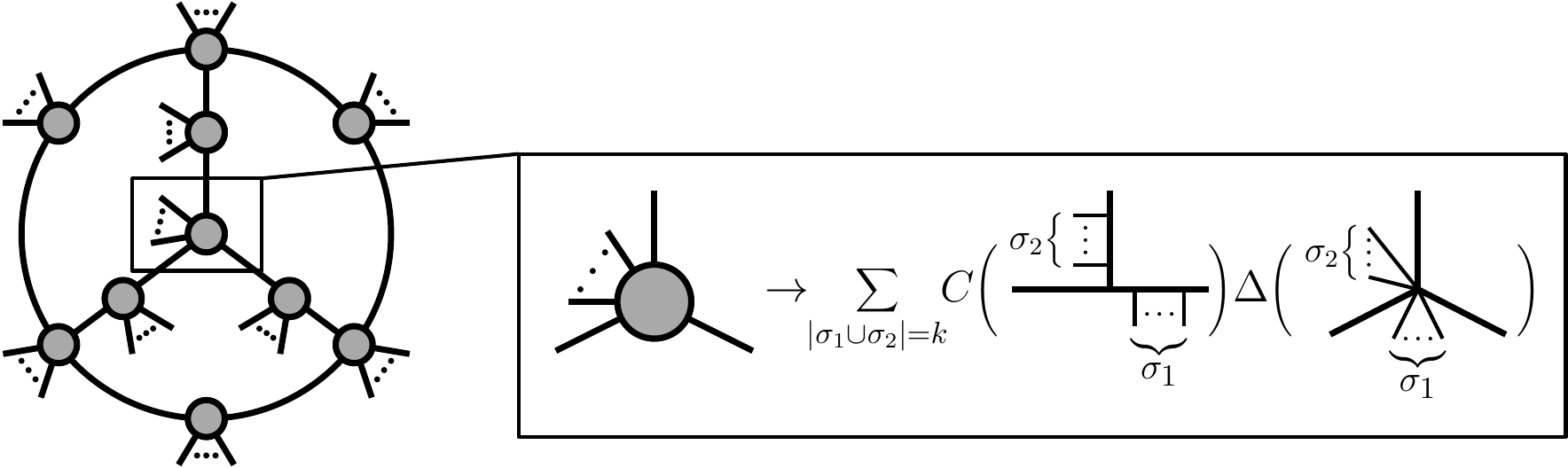}
   \caption{\small Inserting the DDM tree basis into the three-loop coloured cut of topology~$(h)$. The inset treats the case of three loop-dependent legs in an analogous way to the inset~(b) of \fig{fig:color2looppuretopo}.}
   \label{fig:color3loop8topo}
\end{figure}

At three loops, there are eight basic topologies of coloured cuts,
they are shown in \fig{fig:topo3all}.
These involve vertices with up to six loop edges.
In all cases the colour factors are generated by choosing
two loop edges to fix and DDM-decompose,
which gives colour-ordered orderings of the irreducible numerator
with permutations of the external legs and the unfixed loop edges.
In this way, the case of two loop edges has no ambiguity,
as at one and two loops.
Starting from the vertices with three loop legs,
different choices of the two fixed loop edges
produce different colour orderings and colour factors.
For example, the central vertex of topology~$(h)$
can be treated in three different ways,
all of which correspond to some rotation of \fig{fig:color3loop8topo}.

\begin{figure}[t]
   \centering
   \includegraphics[width=0.91\textwidth]{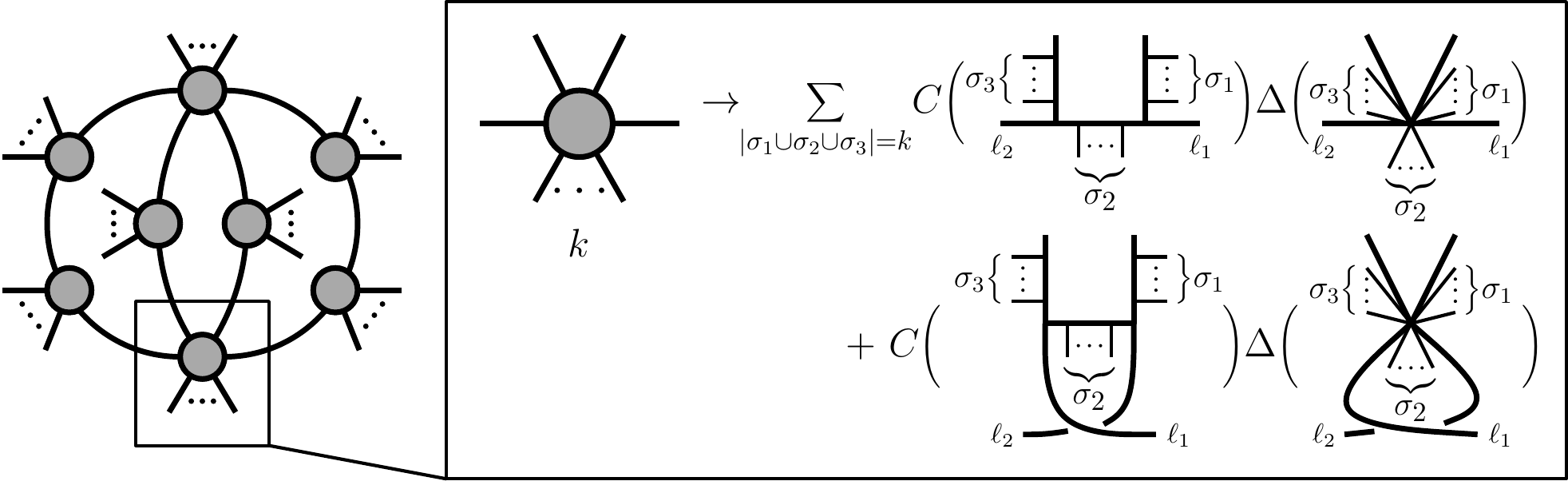}
   \caption{\small Inserting the DDM tree basis into the three-loop coloured cut of topology~$(d)$. The inset treats the case of four loop-dependent legs and shows the two permutations of the loop legs labelled $\ell_1$ and $\ell_2$,
which is analogous to \fig{fig:color2loopbutterflytopo}.}
   \label{fig:color3loop4topo}
\end{figure}

It is preferable to pick the fixed loop legs
with regard to the symmetry of the topology. For instance,
topology~$(d)$ contains two vertices with four loop-dependent edges,
and the four branches can be considered equivalent for the unordered graph.
However, the colour decomposition makes better use of the specifics of that topology if the upper vertex is treated by fixing the same two loop branches
that are fixed in the treatment of the lower vertex
shown in \fig{fig:color3loop4topo}.

\begin{figure}[t]
   \centering
   \includegraphics[width=\textwidth]{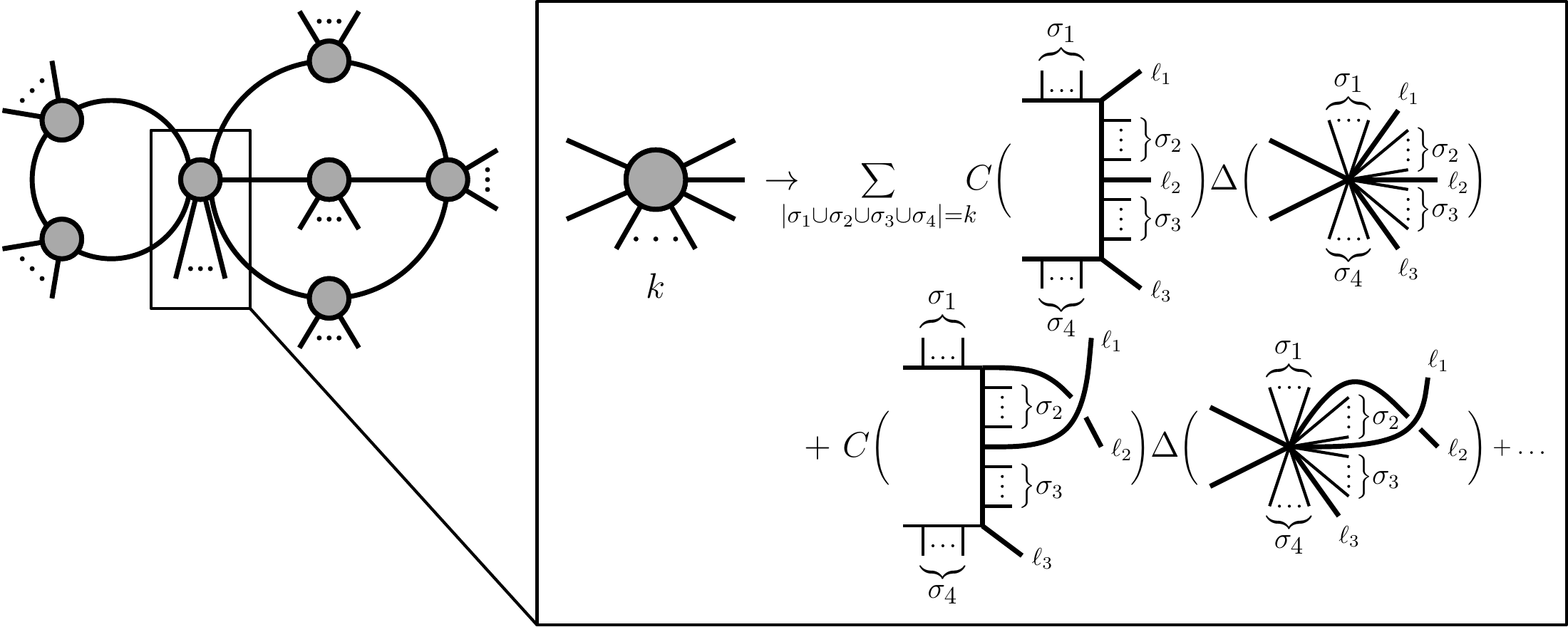}
   \caption{\small Inserting the DDM tree basis into the three-loop coloured cut of topology~$(f)$. The inset treats the case of five loop-dependent legs and shows explicitly two out of 6 permutations of the loop legs labelled $\ell_1$, $\ell_2$ and $\ell_3$.}
   \label{fig:color3loop6topo}
\end{figure}

The vertices with five and six loop-dependent legs
occur only in topologies~$(f)$ and~$(g)$
and are treated in \figs{fig:color3loop6topo}{fig:color3loop7topo}.
The corresponding permutations involve three and four unfixed loop legs,
respectively, and thus generate 6 and 24 ordered topologies.
For simplicity, only the first couple of them are depicted.

\begin{figure}[t]
   \centering
   \includegraphics[width=\textwidth]{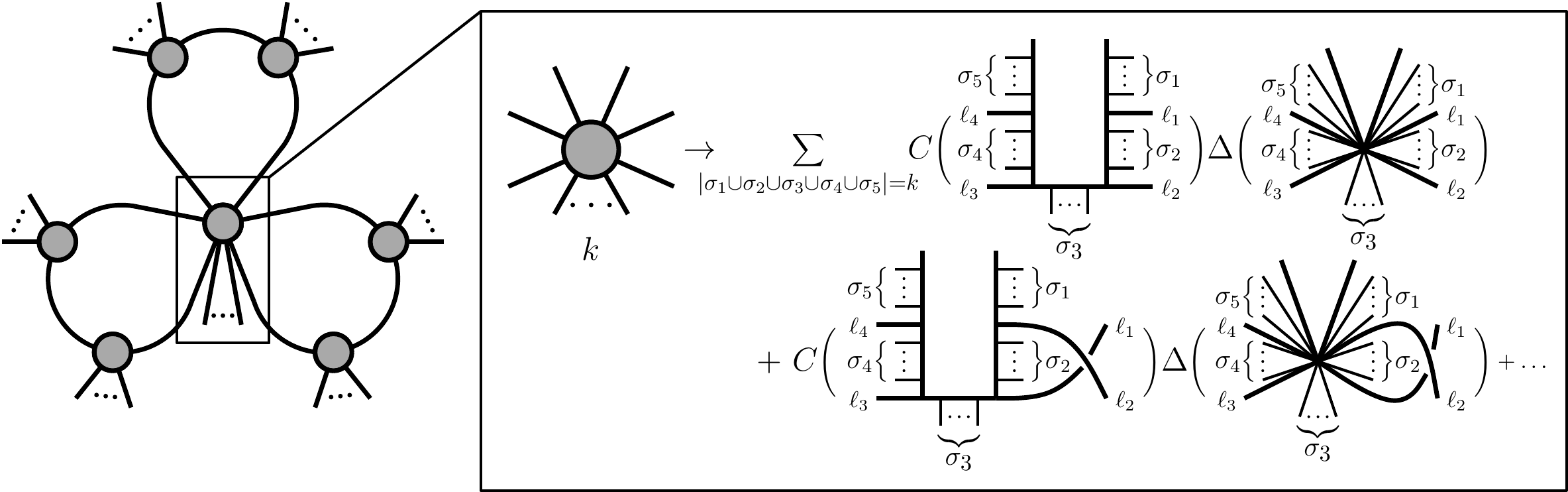}
   \caption{\small Inserting the DDM tree basis into the three-loop coloured cut of topology~$(g)$. The inset treats the case of six loop-dependent legs and shows explicitly two out of 24 permutations of the loop legs labelled $\ell_1$, $\ell_2$, $\ell_3$ and $\ell_4$.}
   \label{fig:color3loop7topo}
\end{figure}

%%%%%%%%%%%%%%%%%%%%%%%%%%%%%%%%%%%%%%%%%%%%%%%%%%%%
\subsubsection{Four-point three-loop example}
\label{sec:4point3loop}
%%%%%%%%%%%%%%%%%%%%%%%%%%%%%%%%%%%%%%%%%%%%%%%%%%%%

As a simple example of the three-loop colour decomposition,
we consider the four-gluon amplitude
in ${\cal N}=4$ supersymmetric Yang-Mills theory,
extensively studied in the literature at the integrand level~\cite{Bern:2007hh,
Bern:2008pv,Bern:2010ue,Bern:2010tq,Bern:2012uf,Bern:2014kca,Bern:2015ple},
as well as recently integrated in \rcite{Henn:2016jdu}.

For convenience, it is commonplace to encode
all the helicity configurations of the external states
inside the overall kinematic prefactor
\be
   {\cal K} = i s_{12} s_{14} A_4^{(0)}(1,2,3,4) .
\label{N4treefactor}
\ee
The tree-level colour-ordered amplitude to which it is proportional
can be understood either as a super-amplitude
or any one of its nonzero MHV components.

The internal-momentum parametrisation of the irreducible numerators
is chosen restricting to irreducible scalar products (ISPs)
purely linear in loop momenta.\footnote{Here and below,
we used the automated Mathematica package \texttt{BasisDet}~\cite{Zhang:2012ce}
to verify irreducibility.}
Moreover, we choose the basis elements in such a way that the resulting
numerators respect the symmetries of their topologies,
which often~\cite{Badger:2015lda} makes their treatment easier.
We find
\vspace{-5pt}
\begin{subequations}
\begin{align} \!\!\!\!\!
\label{N4numeratorsAtoDirr}
   \Delta\!\left(\scalegraph{0.28}{37}{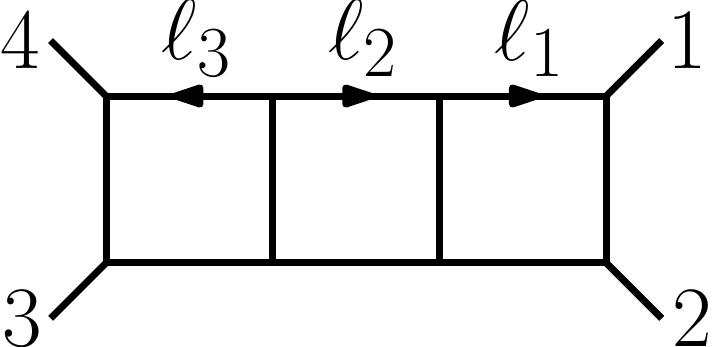}\right) &
 = \Delta\!\left(\scalegraph{0.28}{37}{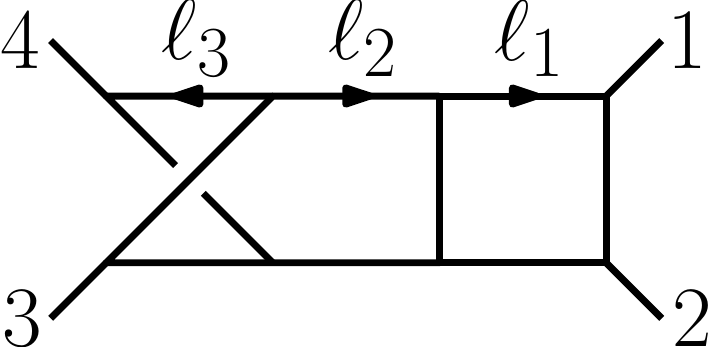}\right)
 = \Delta\!\left(\scalegraph{0.28}{37}{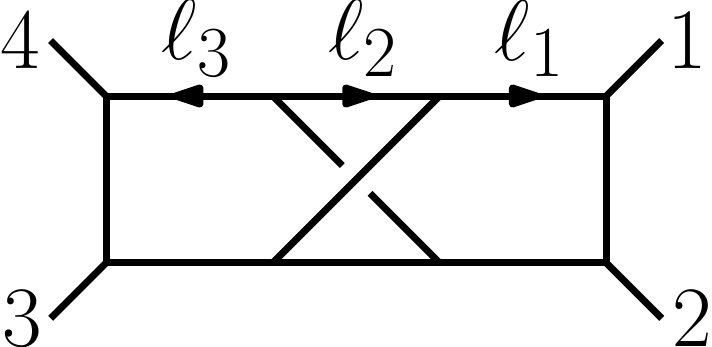}\right)
 = \Delta\!\left(\scalegraph{0.28}{37}{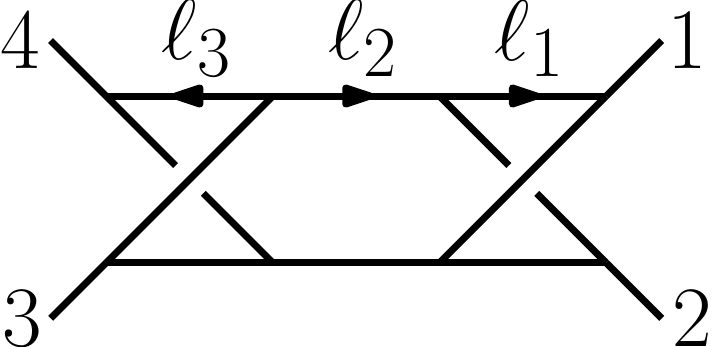}\right)
 = {\cal K} s_{12}^2 , \\
\label{N4numeratorsEtoGirr}
   \Delta\!\left(\:\!\scalegraph{0.28}{65}{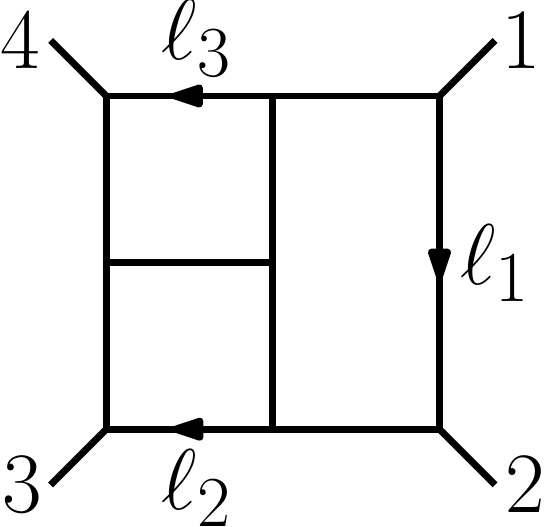}\:\!\right) &
 = \Delta\!\left(\:\!\scalegraph{0.28}{65}{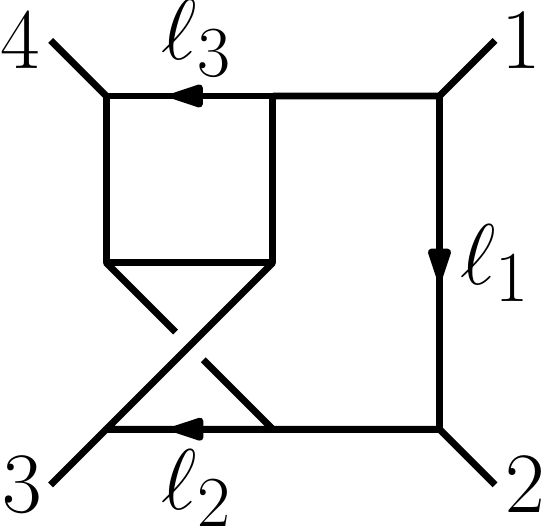}\:\!\right)
 = \Delta\!\left(\:\!\scalegraph{0.28}{62}{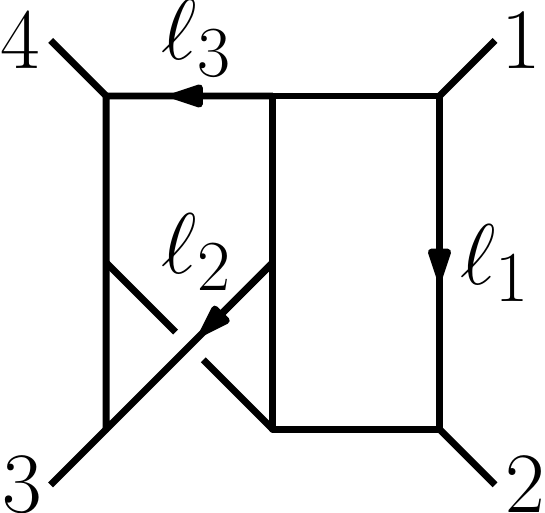}\:\!\right)
 = {\cal K} s_{12} \big(s_{14}+2\ell_1\!\cdot\!(p_1\!+\!p_4) \big) , \!\! \\
\label{N4numeratorsJtoLirr}
   \Delta\!\left(\:\!\scalegraph{0.28}{65}{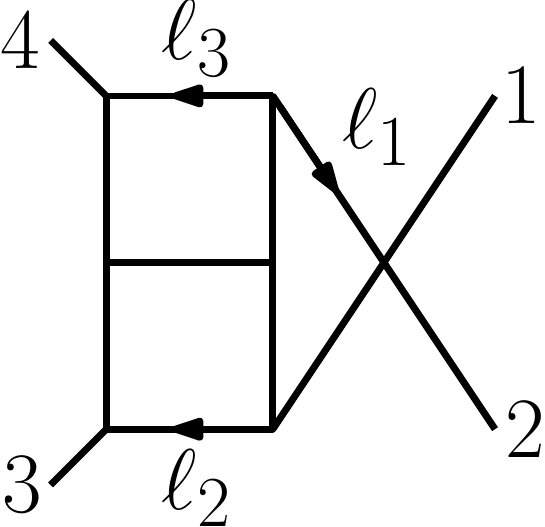}\:\!\right) &
 = \Delta\!\left(\:\!\scalegraph{0.28}{65}{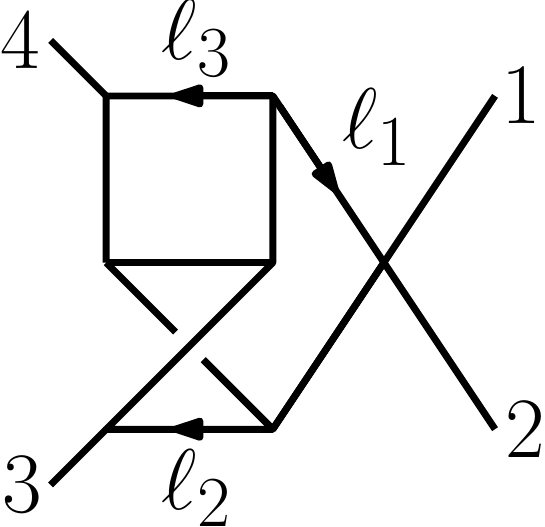}\:\!\right)
 = \Delta\!\left(\:\!\scalegraph{0.28}{62}{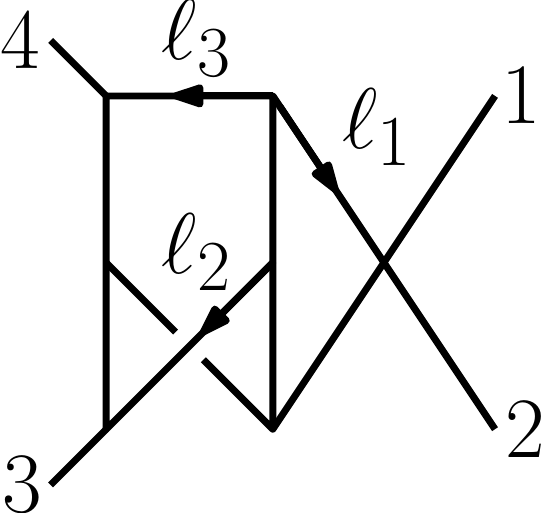}\:\!\right)
 = {\cal K} s_{12} , \\
\label{N4numeratorHirr}
   \Delta\!\left(\:\!\scalegraph{0.28}{62}{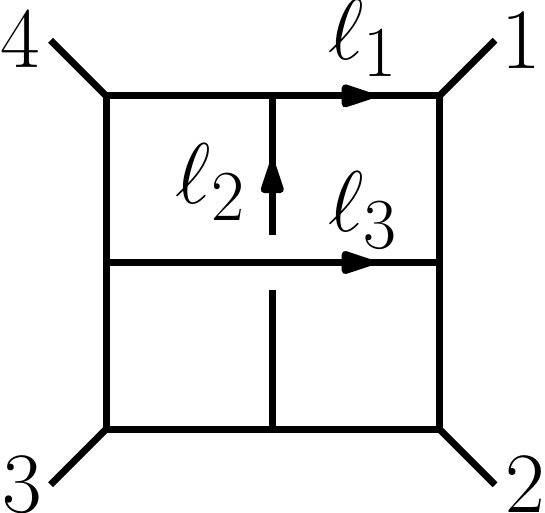}\:\!\right) &
 = {\cal K} \big( s_{12}s_{14}
 - s_{12}\;\!2 \ell_2\!\cdot\!(p_1+p_4)
 - s_{14}\;\!2 \ell_3\!\cdot\!(p_1+p_2) \big) , \\
\label{N4numeratorIirr}
%   \Delta\!\left(\scalegraph{0.28}{60}{graphIi}\right) & \equiv
   \Delta\!\left(\:\!\scalegraph{0.28}{62}{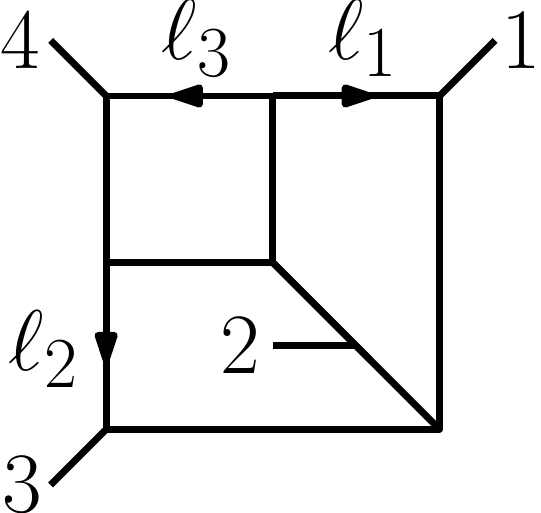}\:\!\right) &
 = {\cal K} \big( s_{12}\;\!2\ell_1\!\cdot\!p_4
 - s_{14}\;\!2\ell_2\!\cdot\!p_4 \big) , \qquad \quad
   \Delta\!\left(\:\!\scalegraph{0.28}{62}{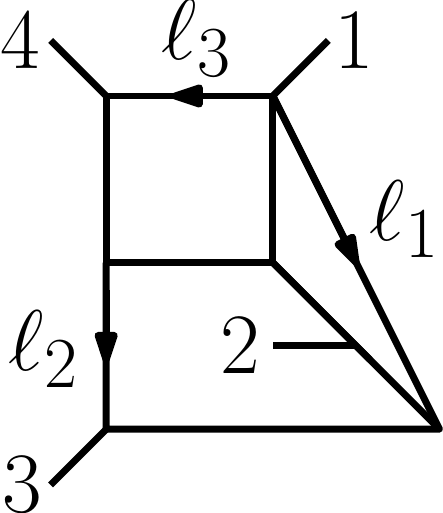}\right)
 = {\cal K} s_{12} , \\
\label{N4numeratorIirr1}
   \Delta\!\left(\scalegraph{0.28}{70}{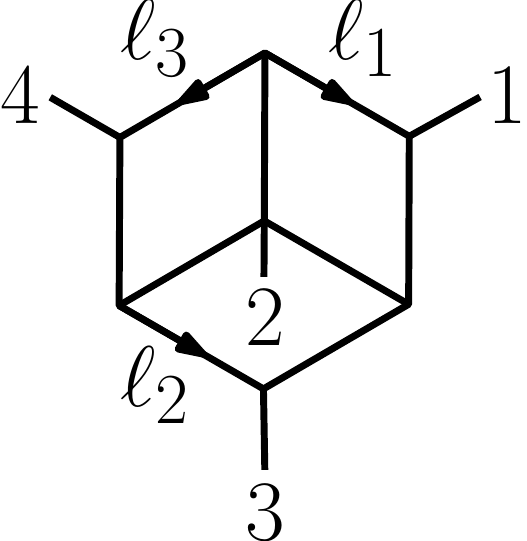}\right) &
 = {\cal K} (2s_{14}-s_{12}-s_{13})/3 ,
\end{align}
\label{N4numeratorsIrr}%
\end{subequations}
where we have shown only non-vanishing topologies.
Note that the numerators~\eqref{N4numeratorsEtoGirr} contain
a reducible scalar product $(\ell_1\!\cdot\!p_1)=\ell_1^2-(\ell_1-p_1)^2$,
but it is combined together with the irreducible $(\ell_1\!\cdot\!p_4)$
into an ISP that respects the flip symmetry
$(\ell_1\!\cdot\!(p_1+p_4)) = (-\ell_1\!\cdot\!(p_2+p_3))$
of the first and the second topologies in \eqn{N4numeratorsEtoGirr}.

\iffalse
% Comments on irredicuble scalar products
The three topologies~\eqref{N4numeratorsEtoGirr}
share the ISP set
$\{(\ell_1\!\cdot\!p_4),(\ell_2\!\cdot\!p_1),(\ell_2\!\cdot\!p_2),
                        (\ell_3\!\cdot\!p_1),(\ell_3\!\cdot\!p_2)\}$.
The ISP set for topology~\eqref{N4numeratorHirr} includes
$\{(\ell_2\!\cdot\!p_1),(\ell_2\!\cdot\!p_4),
   (\ell_3\!\cdot\!p_1),(\ell_3\!\cdot\!p_2)\}$
and either $(\ell_2\!\cdot\!p_2) $ or $(\ell_3\!\cdot\!p_4)$,
while the one for the topology~\eqref{N4numeratorIirr}
$\{(\ell_1\!\cdot\!p_4),(\ell_2\!\cdot\!p_4),
   (\ell_3\!\cdot\!p_1),(\ell_3\!\cdot\!p_2)\}$
and either $(\ell_1\!\cdot\!p_2) $ or $(\ell_2\!\cdot\!p_1)$.
\fi

The full amplitude for arbitrary helicity states can now be constructed
from the irreducible numerators~\eqref{N4numeratorsIrr},
\beal
\label{N4point3loopIrr} \!\!\!\!\!
   \cA^{(3)}_{{\cal N}=4}(1,2,3,4)
    = i g^8
      \sum_{\sigma \in S_4} \sigma \circ
      I\Bigg[ \frac{1}{8}
              C\!\left(\:\!\scalegraph{0.28}{62}{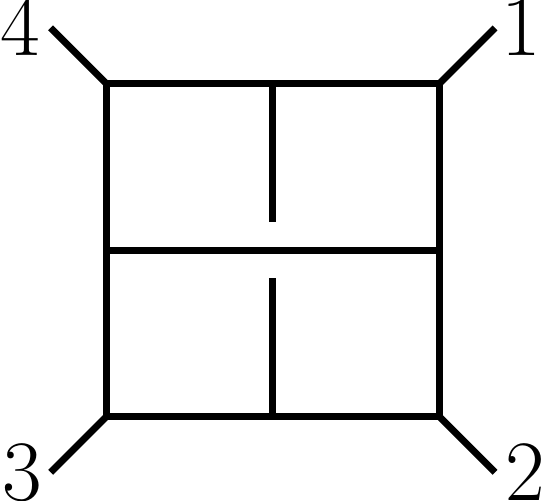}\:\!\right)
              \Delta\!\left(\:\!\scalegraph{0.28}{62}{graphHi}\:\!\right)
              \qquad \qquad \qquad \quad~& \\
         \!+\,\frac{1}{4}
              C\!\left(\:\!\scalegraph{0.28}{38}{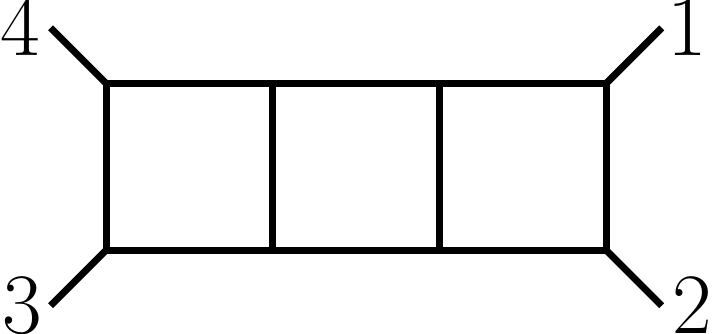}\:\!\right)
              \Delta\!\left(\scalegraph{0.28}{37}{graphAi}\right)
         \! + \frac{1}{4}
              C\!\left(\:\!\scalegraph{0.28}{38}{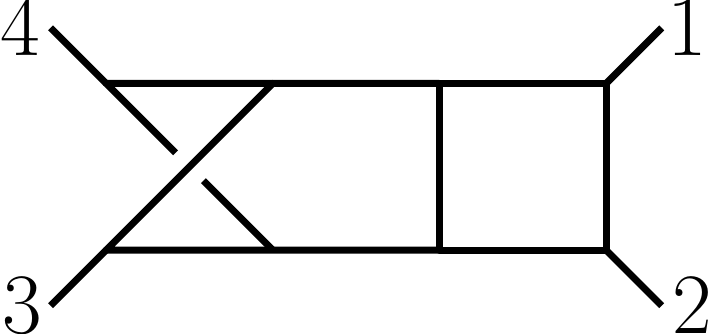}\:\!\right)
              \Delta\!\left(\scalegraph{0.28}{37}{graphBi}\right) & \!\!\! \\
         \!+\,\frac{1}{8}
              C\!\left(\:\!\scalegraph{0.28}{38}{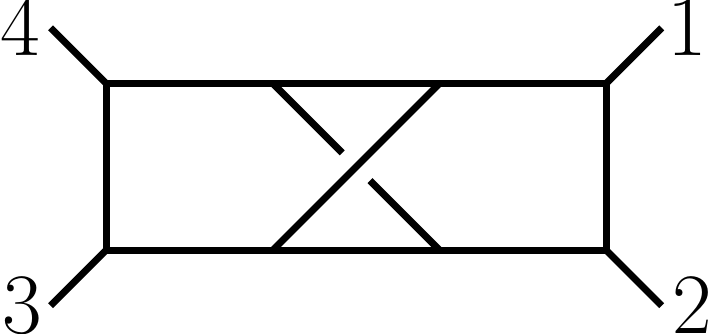}\:\!\right)
              \Delta\!\left(\scalegraph{0.28}{37}{graphCi}\right)
         \! + \frac{1}{16}
              C\!\left(\:\!\scalegraph{0.28}{38}{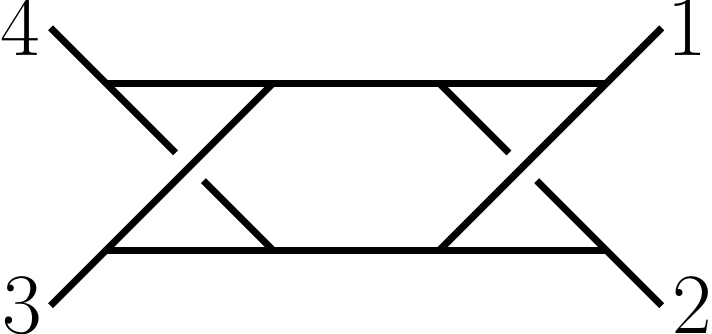}\:\!\right)
              \Delta\!\left(\scalegraph{0.28}{37}{graphDi}\right) & \!\!\! \\
         \!+\,\frac{1}{2}
              C\!\left(\:\!\scalegraph{0.28}{62}{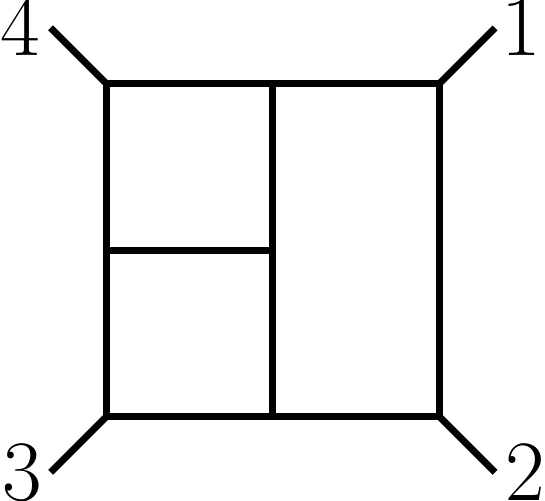}\:\!\right)
              \Bigg\{
              \Delta\!\left(\:\!\scalegraph{0.28}{65}{graphEi}\:\!\right)
         \! + \Delta\!\left(\:\!\scalegraph{0.28}{65}{graphJi}\:\!\right) &
              \Bigg\} \!\!\!\!\! \\
         \!+\,\frac{1}{2}
              C\!\left(\:\!\scalegraph{0.28}{62}{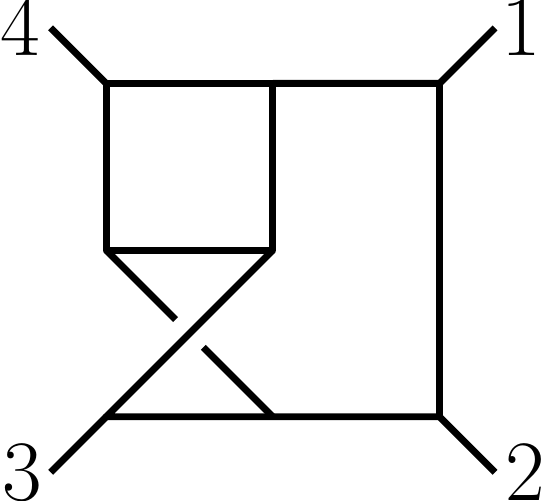}\:\!\right)
              \Bigg\{
              \Delta\!\left(\:\!\scalegraph{0.28}{65}{graphFi}\:\!\right)
         \! + \Delta\!\left(\:\!\scalegraph{0.28}{65}{graphKi}\:\!\right) &
              \Bigg\} \!\!\!\!\! \\
         \!+\,C\!\left(\:\!\scalegraph{0.28}{62}{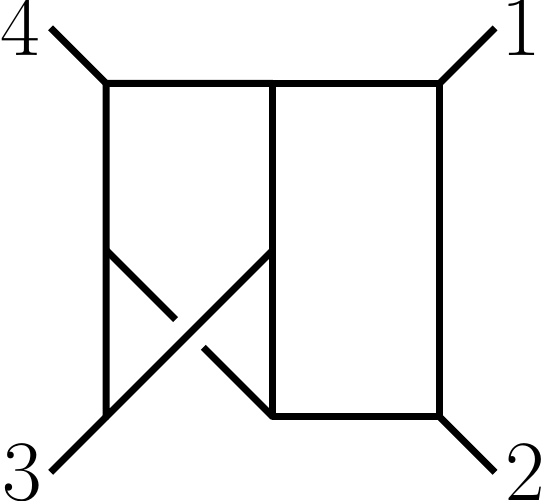}\:\!\right)
              \Bigg\{
              \Delta\!\left(\:\!\scalegraph{0.28}{62}{graphGi}\:\!\right)
         \! + \Delta\!\left(\:\!\scalegraph{0.28}{62}{graphLi}\:\!\right) &
              \Bigg\} \!\!\!\!\! \\
         \!+\,C\!\left(\:\!\scalegraph{0.28}{62}{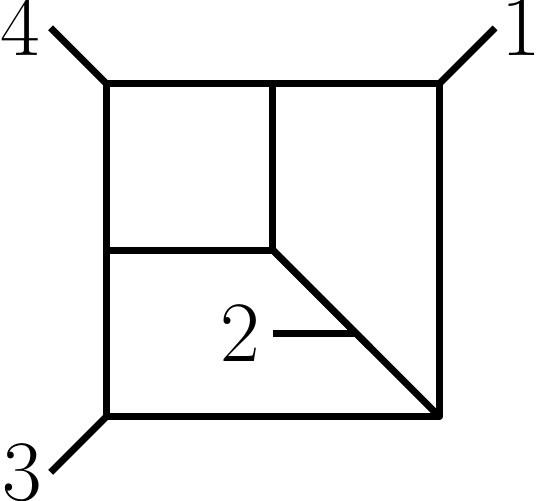}\:\!\right)
              \Bigg\{
              \frac{1}{2}
              \Delta\!\left(\:\!\scalegraph{0.28}{62}{graphIalti}\:\!\right)
         \! + \Delta\!\left(\:\!\scalegraph{0.28}{62}{graphI2i}\right)
         \! + \frac{1}{3}
              \Delta\!\left(\scalegraph{0.28}{70}{graphI1i}\right) &
              \Bigg\} \Bigg] , \!\!\!\!\!\!\!\!
\eeal
where the numerical prefactors are there just to remove overcounting
caused by the full overall permutation sum, except for the fifth diagram
which has a genuine symmetry factor $2$. All colour factors are cubic,
both for the top-level and lower-level numerators.
The most subtle colour decomposition here occurs in the last line.
The DDM-based decomposition for
the colour-dressed version of the topology~\eqref{N4numeratorIirr1} is
\beal \!\!\!
   \tilde{\Delta}\!\left(\scalegraph{0.28}{70}{graphI1i}\right)
    = C\!\left(\:\!\scalegraph{0.28}{62}{graphIaltc}\:\!\right)
      \Delta\!\left(\scalegraph{0.28}{70}{graphI1i}\right)
    + C\!\left(\:\!\scalegraph{0.28}{62}{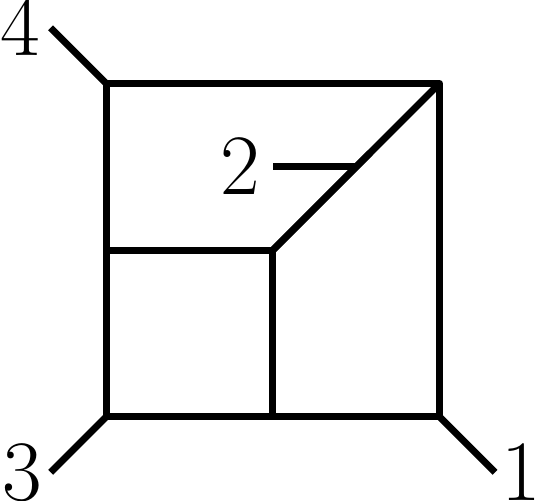}\:\!\right)
      \Delta\!\left(\scalegraph{0.28}{70}{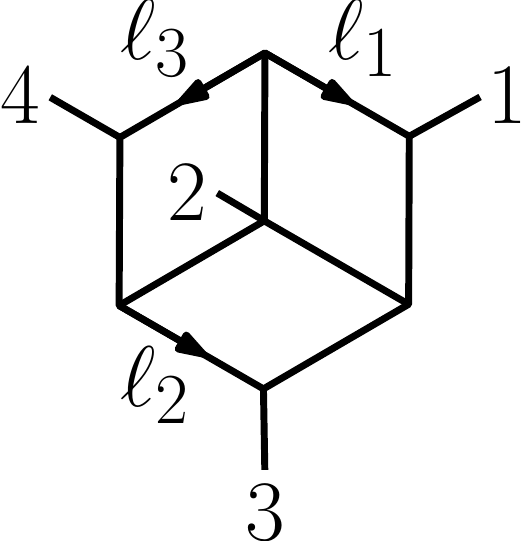}\right) ,
\label{deltaI1DDM}
\eeal
where we ``stretched'' the numerator diagram by its top and right
internal ``spokes''
analogously to the inset~(b) of \fig{fig:color2looppuretopo}.
We could have just as well picked two other pairs of internal propagators.
Regardless, the minimal sum should include \eqn{deltaI1DDM}
and its three other non-equivalent permutations,
corresponding to legs 2, 1, 3 and 4 chosen in the middle of the diagram.
The overall permutation sum in \eqn{N4point3loopIrr} is therefore overcomplete
and effectively symmetrises
over the three choices of internal propagators to ``stretch'' by
in order to decompose the colour-dressed numerator~\eqref{deltaI1DDM}.

In \app{app:4point3loop} we explicitly map our
colour decomposition~\eqref{N4point3loopIrr} to the cubic-graph decomposition
found in \rcites{Bern:2007hh,Bern:2008pv,Bern:2010tq}.

%%%%%%%%%%%%%%%%%%%%%%%%%%%%%%%%%%%%%%%%%%%%%%%%%%%%
\section{Summary and outlook}
%%%%%%%%%%%%%%%%%%%%%%%%%%%%%%%%%%%%%%%%%%%%%%%%%%%%

We have discussed a method to promote tree-level colour decompositions to loop level in the framework of integrand reduction
via unitarity cuts. Effectively, one takes a colour decomposition of a
coloured cut and promotes it to a colour decomposition of the
associated irreducible numerators. This can be achieved with any given
colour decomposition, but is most powerful when applied to one which
is KK-independent. We have shown that colour-ordered irreducible
numerators can be chosen to inherit KK relations of their associated
cuts and demonstrated how this can be used to simplify amplitude calculations
in gauge theories with various degrees of supersymmetry. As the
irreducible numerators live under an integral sign,
different momentum reroutings of the same topology are equivalent,
and this can used to cancel the symmetry factors, at least at one and two loops.
At one loop, this allows the method
to recover the DDM one-loop decomposition~\cite{DelDuca:1999rs}.
At two loops, we have supplemented
the five-point colour decomposition of \rcite{Badger:2015lda}
by the maximal and next-to-maximal topologies
that correspond to vanishing cuts for the all positive external helicities
but should be included in other helicity configurations.
As a three-loop example, we have also considered the four-point amplitude
in maximally supersymmetry Yang-Mills theory~\cite{Bern:2007hh}.
More generally, we have shown how the DDM-based colour decomposition applies
to any type of three-loop cut topology.

In this note, we considered the multi-loop integrand construction using irreducible numerators.
We have assumed that the numerator KK relations hold at the integrand level,
which can be helped by a loop-momentum parametrisation of a numerator
that is tailored to its symmetries.
However, this assumption can be removed provided that the integrated amplitude
stays the same or, equivalently, it satisfies all the unitarity cuts.
Hence there should be a way to use the presented colour decomposition
in absence of off-shell KK relations,
though self-consistency of the hierarchy subtraction
may be more difficult to achieve in this case.
For instance, this may be relevant in the framework of non-planar on-shell diagrams~\cite{Franco:2013nwa,Chen:2014ara,Arkani-Hamed:2014bca,Franco:2015rma,Chen:2015qna,Bern:2015ple,Bourjaily:2016mnp},
where the requirement of the irreducibility is replaced by a connection to the amplituhedron structure~\cite{ArkaniHamed:2012nw,Arkani-Hamed:2013jha}.
It would be interesting to use the presented colour decomposition
in this setting.

Moreover, in future work we hope to further explore this technique in the
context of QCD amplitudes with quarks~\cite{Mangano:1988kk,Mangano:1990by,Maltoni:2002mq,Ita:2011ar}.
Indeed, the recent studies of the tree-level KK relations
for multi-quark amplitudes~\cite{Reuschle:2013qna,Schuster:2013aya,Melia:2013bta,Melia:2013epa,Melia:2014oza},
have led to a proper colour decomposition~\cite{Johansson:2015oia,Melia:2015ika}
of a quark-gluon amplitude into KK-independent basis of partial amplitudes~\cite{Melia:2013bta,Melia:2013epa},
analogous in this respect to the DDM formula~\cite{DelDuca:1999rs}.
In this way, the presented method will be applicable to a greater variety of
phenomenologically pertinent amplitude calculations.
\nocite{Ohl:1995kr}

%%%%%%%%%%%%%%%%%%%%%%%%%%%%%%%%%%%%%%%%%%%%%%%%%%%%
\begin{acknowledgments}

We are very grateful to
Simon Badger, Fernando Febres-Cordero, Harald Ita, Gustav Mogull and Donal O'Connell
for enlightening discussions and for collaboration on related topics.
We thank Harald Ita and Donal O'Connell for useful comments on the draft,
as well as Samuel Abreu, Matthieu Jaquier, Tiziano Peraro and Christopher Schwan for helpful discussions.
AO is supported in part by the Marie Curie FP7 grant 631370.
BP is supported by the Alexander von Humboldt Foundation, in the
framework of the Sofja Kovalevskaja Award 2014, endowed by the German
Federal Ministry of Education and Research.
AO thanks Nordita for hospitality during
the programme ``Aspects of Amplitudes.''

\end{acknowledgments}
%%%%%%%%%%%%%%%%%%%%%%%%%%%%%%%%%%%%%%%%%%%%%%%%%%%%

%%%%%%%%%%%%%%%%%%%%%%%%%%%%%%%%%%%%%%%%%%%%%%%%%%%%
\appendix

%%%%%%%%%%%%%%%%%%%%%%%%%%%%%%%%%%%%%%%%%%%%%%%%%%%%
\section{Bubble-insertion subtlety}
\label{app:subtlety}
%%%%%%%%%%%%%%%%%%%%%%%%%%%%%%%%%%%%%%%%%%%%%%%%%%%%

In \eqn{5pointPureYMDecomposition} we used
the factorisation limit of the following relation:
\beal
   \frac{1}{2} \tilde{\Delta}\bigg(\eqnDelta{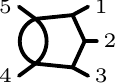}\!\!\!\:\bigg)
 = C\bigg(\eqnDelta{delta431i}\!\!\bigg)
   \Delta\bigg(\eqnDelta{delta411i}\!\!\!\:\bigg)
 + C\bigg(\!\!\eqnDelta{delta422NPi}\!\!\bigg)
   \Delta\bigg(\eqnDelta{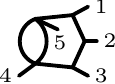}\!\!\!\:\bigg) ,
\eeal
which is assumed to hold under the integration sign
similarly to the result of \eqn{eq:trianglebubbleexpansion}.
This implicitly relies on the KK relations for the colour-ordered numerators
and corresponding cuts, such as
\beal
   0 &
 = \text{Cut}\bigg(\!\!\eqnDelta{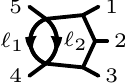}\!\!\!\:\bigg)
 + \text{Cut}\bigg(\!\!\eqnDelta{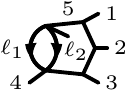}\!\!\!\:\bigg)
 + \text{Cut}\bigg(\!\!\eqnDelta{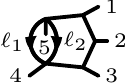}\!\!\!\:\bigg) \\ &
 = \Delta\bigg(\!\!\eqnDelta{delta411l}\!\!\!\:\bigg)
 + \Delta\bigg(\!\!\eqnDelta{delta411UNP2l}\!\!\!\:\bigg)
 + \Delta\bigg(\!\!\eqnDelta{delta411UNP1l}\!\!\!\:\bigg) \\ & \quad
 + \frac{1}{(\ell_1+\ell_2)^2}
   \bigg\{
   \Delta\bigg(\eqnDelta{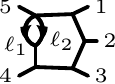}
              +\eqnDelta{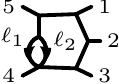}\!\!\!\:\bigg)
 + \Delta\bigg(\eqnDelta{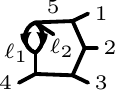}
              +\eqnDelta{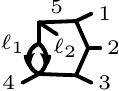}\!\!\!\:\bigg)
 + \Delta\bigg(\eqnDelta{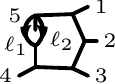}\!\!\!\:\bigg)
   \bigg\} ,
\label{KKbubbleinsertion}
\eeal
where all the other higher-level topologies cancel in the usual way.
The subtlety here is that the topologies like
$\eqnDiag{\scalegraph{0.64}{0}{delta511Uc}}$
do not correspond to well-defined cuts.
Instead, we understand the numerator
$\Delta\big(\!\eqnDiag{\scalegraph{0.64}{0}{delta511Uc}}
         +\!\!\eqnDiag{\scalegraph{0.64}{0}{delta511Dc}}\;\!\!\big)$
as a single coefficient function of the $1/(\ell_1+\ell_2)^2$ pole
inside the finite
$\text{Cut}\big(\!\eqnDiag{\scalegraph{0.64}{0}{delta411c}}\;\!\!\big)$.
This prevents us from separating and cancelling the graphs
in the last line of \eqn{KKbubbleinsertion},
as it would normally happen.
Therefore, \eqn{KKbubbleinsertion}
implies not only the KK relation between
$\Delta\big(\!\eqnDiag{\scalegraph{0.64}{0}{delta411c}}\;\!\!\big)$,
$\Delta\big(\!\eqnDiag{\scalegraph{0.64}{0}{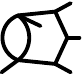}}\;\!\!\big)$
and
$\Delta\big(\!\eqnDiag{\scalegraph{0.64}{0}{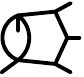}}\;\!\!\big)$
but also an accompanying KK relation of less familiar form:
\be
   \Delta\bigg(\eqnDelta{delta511UNP1l}\!\!\!\:\bigg) =
 - \Delta\bigg(\eqnDelta{delta511U1l}
              +\eqnDelta{delta511D1l}\!\!\!\:\bigg)
 - \Delta\bigg(\eqnDelta{delta511UNP2l}
              +\eqnDelta{delta511DNP3l}\!\!\!\:\bigg) .
\ee

%%%%%%%%%%%%%%%%%%%%%%%%%%%%%%%%%%%%%%%%%%%%%%%%%%%%
\section{Check of four-point three-loop example}
\label{app:4point3loop}
%%%%%%%%%%%%%%%%%%%%%%%%%%%%%%%%%%%%%%%%%%%%%%%%%%%%

Here let us show how our amplitude decomposition~\eqref{N4point3loopIrr}
relates to the cubic-graph decomposition found
in \rcites{Bern:2007hh,Bern:2008pv,Bern:2010tq}:
\begin{align}
\label{N4point3loop}
   \cA^{(3)}_{{\cal N}=4}(1,2,3,4)
    = i g^8 {\cal K}
      \sum_{\sigma \in S_4} \sigma \circ
      I\Bigg[ \frac{1}{4}
              \tilde{N}\!\left(\scalegraph{0.28}{37}{graphAi}\right)
              \qquad \qquad \qquad \quad & \\
         \! + \frac{1}{4}
              \tilde{N}\!\left(\scalegraph{0.28}{37}{graphBi}\right)
         \! + \frac{1}{8}
              \tilde{N}\!\left(\scalegraph{0.28}{37}{graphCi}\right)
         \! + \frac{1}{16}
              \tilde{N}\!\left(\scalegraph{0.28}{37}{graphDi}\right) &
         \! + \frac{1}{2}
              \tilde{N}\!\left(\:\!\scalegraph{0.28}{65}{graphEi}\:\!\right) \nn \\
         \! + \frac{1}{2}
              \tilde{N}\!\left(\:\!\scalegraph{0.28}{65}{graphFi}\:\!\right)
         \! + \tilde{N}\!\left(\:\!\scalegraph{0.28}{62}{graphGi}\:\!\right)
         \! + \frac{1}{8}
              \tilde{N}\!\left(\:\!\scalegraph{0.28}{62}{graphHi}\:\!\right) &
         \! + \frac{1}{2}
              \tilde{N}\!\left(\:\!\scalegraph{0.28}{62}{graphIalti}\:\!\right) \!
      \Bigg] , \nn
\end{align}
where the purely trivalent colour factors are implicit,
and the kinematic parts of the numerators are given by
\vspace{-15pt}
\begin{subequations}
\begin{align} \label{N4numeratorsAtoD} \\ \nn
   N\!\left(\scalegraph{0.28}{37}{graphAi}\right) &
 = N\!\left(\scalegraph{0.28}{37}{graphBi}\right)
 = N\!\left(\scalegraph{0.28}{37}{graphCi}\right)
 = N\!\left(\scalegraph{0.28}{37}{graphDi}\right)
 = s_{12}^2 , \\
\label{N4numeratorsEtoG}
   N\!\left(\:\!\scalegraph{0.28}{65}{graphEi}\:\!\right) &
 = N\!\left(\:\!\scalegraph{0.28}{65}{graphFi}\:\!\right)
 = N\!\left(\:\!\scalegraph{0.28}{62}{graphGi}\:\!\right)
 = s_{12} (\ell_1+p_1+p_4)^2 , \\
\label{N4numeratorH}
   N\!\left(\:\!\scalegraph{0.28}{62}{graphHi}\:\!\right) &
 = s_{12}s_{14}
 - s_{12}\;\!2 \ell_2\!\cdot\!(p_1+p_4)
 - s_{14}\;\!2 \ell_3\!\cdot\!(p_1+p_2) , \\
\label{N4numeratorI}
   N\!\left(\:\!\scalegraph{0.28}{62}{graphIalti}\:\!\right) &
 = s_{12} (\ell_1+p_4)^2
 - s_{14} (\ell_2+p_4)^2
 - (s_{12}-s_{14})(\ell_1+\ell_2+p_4)^2/3 .
\end{align}
\label{N4numerators}%
\end{subequations}

Here the numerators~\eqref{N4numeratorsAtoD} and~\eqref{N4numeratorH}
equal already their irreducible
counterparts~\eqref{N4numeratorsAtoDirr} and~\eqref{N4numeratorHirr}
(up to the prefactor ${\cal K}$).
The numerators~\eqref{N4numeratorsEtoG} are naturally combined
from their irreducible counterparts~\eqref{N4numeratorsAtoDirr}
and related lower-level topologies~\eqref{N4numeratorsJtoLirr}, \ie
\be
 {\cal K}N\!\left(\:\!\scalegraph{0.28}{65}{graphEi}\:\!\right)
  = \Delta\!\left(\:\!\scalegraph{0.28}{65}{graphEi}\:\!\right)
  + \ell_1^2 \Delta\!\left(\:\!\scalegraph{0.28}{65}{graphJi}\:\!\right) .
\ee

Finally, to obtain the numerator~\eqref{N4numeratorI},
we can use the diagonal-flip antisymmetry of the colour factor
in the last line of \eqn{N4point3loopIrr},
\begin{align}
\label{N4point3loopIpart}
      \sum_{\sigma \in S_4} \sigma \circ
      C\!\left(\:\!\scalegraph{0.28}{62}{graphIaltc}\:\!\right)
      I\Bigg[
      \frac{1}{2}
      \Delta\!\left(\:\!\scalegraph{0.28}{62}{graphIalti}\:\!\right)
 \! + \Delta\!&\left(\:\!\scalegraph{0.28}{62}{graphI2i}\:\!\right)
 \! + \frac{1}{3} \Delta\!\left(\scalegraph{0.28}{70}{graphI1i}\right) \!
      \Bigg] \\
    = \sum_{\sigma \in S_4} \sigma \circ
      C\!\left(\:\!\scalegraph{0.28}{62}{graphIaltc}\:\!\right)
      I\Bigg[
      \frac{1}{2}
      \Delta\!\left(\:\!\scalegraph{0.28}{62}{graphIalti}\:\!\right) &
 \! + \frac{1}{2}
      \Delta\!\left(\:\!\scalegraph{0.28}{62}{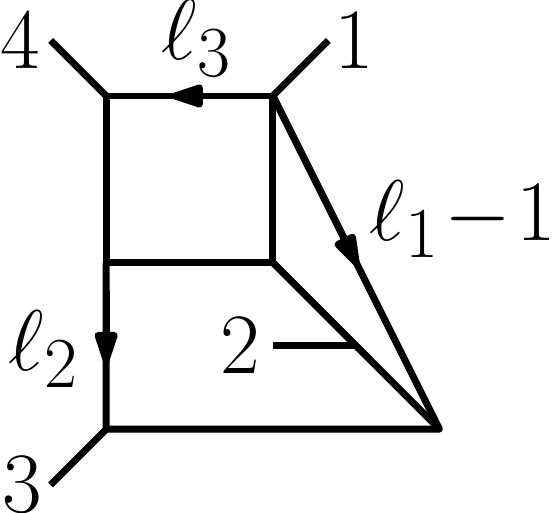}\!\!\right)
 \! - \frac{1}{2}
      \Delta\!\left(\:\!\scalegraph{0.28}{56}{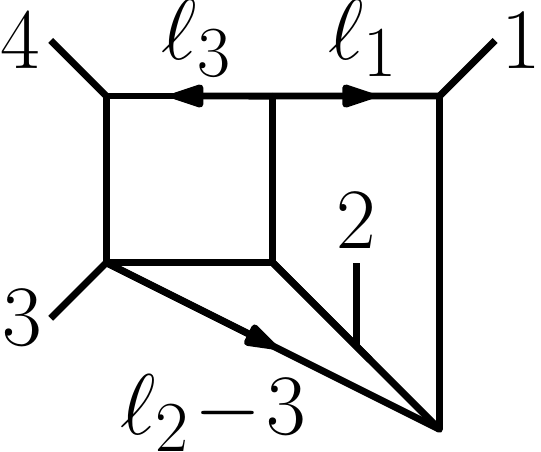}\:\!\right)
       \nn \\ &
 \! + \frac{1}{6}
      \Delta\!\left(\scalegraph{0.28}{70}{graphI1i}\right)
 \! - \frac{1}{6}
      \Delta\!\left(\scalegraph{0.28}{70}{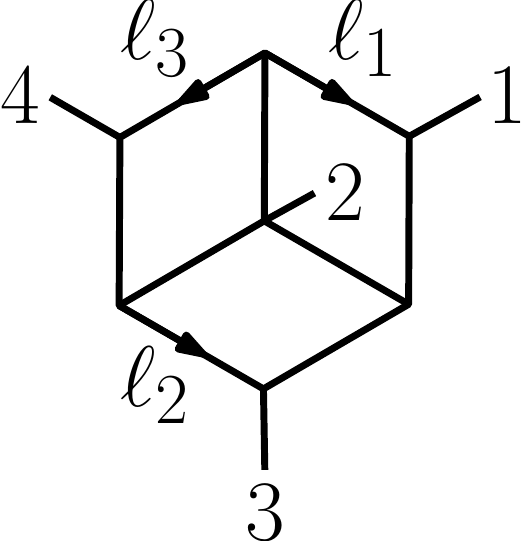}\right) \!
      \Bigg] . \nn
\end{align}
Taking into account that our integration measure~\eqref{IntMeasure}
includes topology-specific propagators, we retrieve precisely
\beal
   {\cal K}N\!\left(\:\!\scalegraph{0.28}{62}{graphIalti}\:\!\right)
    = \Delta\!\left(\:\!\scalegraph{0.28}{62}{graphIalti}\:\!\right)
 \! + \ell_1^2
      \Delta\!\left(\:\!\scalegraph{0.28}{62}{graphIil1reduced}\!\!\right)
 \! - \ell_2^2
      \Delta\!\left(\:\!\scalegraph{0.28}{56}{graphIil2reduced}\:\!\right) & \\
 \! + \frac{1}{3} (\ell_1+\ell_2+p_4)^2
      \bigg\{
      \Delta\!\left(\scalegraph{0.28}{70}{graphI1i}\right)
 \! - \Delta\!\left(\scalegraph{0.28}{70}{graphI1i4213}\right) &
    \bigg\} .
% \\ = {\cal K}
%      \Big\{
%      s_{12} \big( 2\ell_1\!\cdot\!p_4 + \ell_1^2 \big)
%    - s_{14} \big( 2\ell_2\!\cdot\!p_4 + \ell_2^2 \big)
%    + \frac{1}{9} (\ell_1+\ell_2+p_4)^2
%      \big( 3s_{14} - 3s_{12} \big) &
%      \Big\} .
\eeal
In this way,
we map to our irreducible-numerator decomposition~\eqref{N4point3loopIrr}
to the cubic-graph decomposition~\eqref{N4point3loop}.
It should be noted that one can in principle choose irreducible monomials
in such a way that only the top-level numerators~\eqref{N4numerators}
are left non-vanishing.
In \sec{sec:4point3loop} we intentionally restricted ourselves
to the parametrisation at most linear in loop momenta
in order to demonstrate nontrivial elements
of the DDM-based colour decomposition.

%%%%%%%%%%%%%%%%%%%%%%%%%%%%%%%%%%%%%%%%%%%%%%%%%%%%
\bibliographystyle{JHEP}
\bibliography{references}

\end{document}